\documentclass[usenatbib,onecolumn]{mn2e}
\usepackage{epsfig}
\usepackage{color}

\newcommand{\vect}[1]{\mathbf{#1}}
\newcommand{\chng}[1]{#1}
\title[Softening as Smoothing]{Gravitational softening as a smoothing
operation}

\author[J. E. Barnes]{Joshua E. Barnes\\
Institute of Astronomy, University of Hawaii, 2680 Woodlawn Drive,
Honolulu, HI 96822, USA}

\begin{document}

\maketitle

\begin{abstract}
In self-consistent $N$-body simulations of collisionless systems,
gravitational interactions are modified on small scales to remove
singularities and simplify the task of numerically integrating the
equations of motion.
This `gravitational softening' is sometimes presented as \chng{an}
ad-hoc \chng{departure from} Newtonian gravity.
However, softening can also be described as a smoothing operation
applied to the mass distribution; the gravitational potential
\chng{and} the smoothed density \chng{obey} Poisson's
equation \chng{precisely}.
While `softening' and `smoothing' are mathematically equivalent
\chng{descriptions}, the latter has some advantages.
For example, the smoothing description \chng{suggests a way} to set up
$N$-body initial conditions in almost perfect dynamical equilibrium.
\end{abstract}


\section{INTRODUCTION}

The evolution of a collisionless self-gravitating system is described
by two coupled equations: the Vlasov equation,
\begin{equation}
  \frac{\partial f}{\partial t} +
    \vect{v} \cdot \frac{\partial f}{\partial \vect{r}} -
      \nabla \Phi \cdot \frac{\partial f}{\partial \vect{v}} = 0 \, ,
  \label{eq:vlasov}
\end{equation}
where $f = f(\vect{r}, \vect{v}, t)$ is the one-particle distribution
function and $\Phi(\vect{r}, t)$ is the gravitational potential, and
Poisson's equation,
\begin{equation}
  \nabla^2 \Phi = 4 \pi G \rho = 4 \pi G \int d\vect{v} \, f \, .
  \label{eq:poisson}
\end{equation}
$N$-body simulations use a Monte-Carlo method to solve these equations.
The distribution function is represented by a collection of $N$
particles \citep{K67}:
\begin{equation}
  f(\vect{r}, \vect{v}, t) =
    \sum_{i = 1}^{N} m_i \,
                     \delta^3(\vect{r} - \vect{r}_i(t)) \,
                     \delta^3(\vect{v} - \vect{v}_i(t)) \, ,
  \label{eq:dfrep}
\end{equation}
where $m_i$, $\vect{r}_i$, and $\vect{v}_i$ are the mass, position,
and velocity of particle $i$.
Over time, particles move along characteristics of (\ref{eq:vlasov});
at each instant, their positions provide the density needed for
(\ref{eq:poisson}).

In many collisionless $N$-body simulations, the equations of motion
actually integrated are
\begin{equation}
  \frac{d \vect{r}_i}{d t} = \vect{v}_i \, ,
  \qquad
  \frac{d \vect{v}_i}{d t} =
    \sum_{j \ne i}^{N} G m_j
      \frac{\vect{r}_j - \vect{r_i}}
           {(|\vect{r}_j - \vect{r}_i|^2 + \epsilon^2)^{3/2}} \, ,
  \label{eq:nbody}
\end{equation}
where $\epsilon$ is the \textit{softening length}.
These equations reduce to the standard Newtonian equations of motion
if $\epsilon = 0$.
The main reason for setting $\chng{\epsilon \ne 0}$ is to suppress the
$1/r$ singularity in the Newtonian potential; this greatly simplifies
the task of \chng{numerically} integrating these equations
\citep[e.g.,][]{D01}. 
By limiting the spatial resolution of the gravitational force,
softening also helps \chng{control} fluctuations \chng{caused by}
sampling the distribution function with finite $N$; however, this
\chng{comes at a price}, since the gravitational field is
\chng{systematically} biased for $\chng{\epsilon \ne 0}$ \citep{M96,
ABLM98, AFLB00}.

Softening is often described as a modification of Newtonian gravity,
with the $1/r$ potential replaced by $1/\sqrt{r^2+\epsilon^2}$.
The latter is proportional to the potential of a \citet{P11} sphere
with scale radius $\epsilon$.
This does \textit{not} imply that particles interact like Plummer
spheres \citep{DI93}; the acceleration of particle $i$ is computed
from the field at the point $\vect{r}_i$ \chng{only}.
But it does imply that softening can also be described as a smoothing
operation \citep[e.g.,][]{HB90}, in which the pointillistic
Monte-Carlo representation of the density field is convolved with
\chng{the} kernel
\begin{equation}
  S(r;\epsilon) = \frac{3}{4 \pi}
    \frac{\epsilon^2}{(r^2 + \epsilon^2)^{5/2}} \, .
  \label{eq:plummer-kernel}
\end{equation}
In effect, the source term for Poisson's equation (\ref{eq:poisson})
is replaced with the \textit{smoothed density}
\begin{equation}
  \rho(\vect{r};\epsilon) \equiv
    \int d\vect{r}' \, \rho(\vect{r}') S(|\vect{r}-\vect{r}'|;\epsilon) =
    \int d\vect{r}' \, \rho(\vect{r}-\vect{r}') S(|\vect{r}'|;\epsilon) \, .
  \label{eq:smooth-rho}
\end{equation}
Formally, (\ref{eq:nbody}) provides a Monte-Carlo solution to the
Vlasov equation (\ref{eq:vlasov}) coupled with
\begin{equation}
  \nabla^2 \Phi = 4 \pi G \rho(\vect{r};\epsilon) =
    4 \pi G \int d\vect{r}' \, \int d\vect{v} \,
      f(\vect{r}',\vect{v},t) S(|\vect{r}-\vect{r}'|;\epsilon) \, .
  \label{eq:smooth-poisson}
\end{equation}
Thus one may argue that a softened $N$-body simulation actually uses
standard Newtonian gravity, as long as \chng{it is clear} that the
mass distribution generating the gravitational field is derived from
the particles via a smoothing process.

Although Plummer softening is widely used in $N$-body
\chng{simulations}, its
effects are incompletely understood.
If the underlying density field is featureless on scales \chng{of
order $\epsilon$}, softening has relatively little effect.
However, $N$-body simulations \chng{are} often used to model systems
with power-law density profiles; for example, \citet{H90} and
\citet[][\chng{hereafter NFW}]{NFW96} models, which have $\rho \propto
r^{-1}$ at small $r$, are widely used as initial conditions.
One purpose of this paper is to examine how softening modifies such
profiles.

Assume that the underlying density profile is spherically symmetric
and centered on the origin: $\rho = \rho(|\vect{r}|)$.
\chng{The integrand in (\ref{eq:smooth-rho}) is unchanged by
rotation about the axis containing the origin and the point
$\vect{r}$, so the integral can be simplified by adopting cylindrical
coordinates $(\varphi, R, z)$, where $\vect{r}$ is located on the
$z$-axis at $z = |\vect{r}|$.
The integral over $\varphi$ is trivial; for Plummer smoothing, the
result is
%%%%%%%%%%%%%%%%%%%%%%%%%%%%%%%%%%%%%%%%%%%%%%%%%%%%%%%%%%%%%%%%%%%%%%%%
\begin{equation}
  \rho(r;\epsilon) =
    \frac{3 \epsilon^2}{2}
    \int_{-\infty}^{\infty} dz
      \int_{0}^{\infty} dR \, R
        \frac{\rho({\textstyle \sqrt{R^2 + z^2}})}
             {(R^2 + (z - r)^2 + \epsilon^2)^{5/2}} =
    \frac{3 \epsilon^2}{2}
    \int_{-\infty}^{\infty} dz
      \int_{0}^{\infty} dR \, R
        \frac{\rho({\textstyle \sqrt{R^2 + (z - r)^2}})}
             {(R^2 + z^2 + \epsilon^2)^{5/2}} \, ,
  \label{eq:plummer-smooth}
\end{equation}
%%% \begin{equation}
%%%   \textstyle
%%%   \rho(r;\epsilon) =
%%%     \frac{3 \epsilon^2}{2}
%%%     {\displaystyle \int_{-\infty}^{\infty}} dz
%%%       {\displaystyle \int_{0}^{\infty}} dR \, R
%%%         \frac{\rho({\scriptstyle \sqrt{R^2 + z^2}})}
%%%              {(R^2 + (z - r)^2 + \epsilon^2)^{5/2}} =
%%%     \frac{3 \epsilon^2}{2}
%%%     {\displaystyle \int_{-\infty}^{\infty}} dz
%%%       {\displaystyle \int_{0}^{\infty}} dR \, R
%%%         \frac{\rho({\scriptstyle \sqrt{R^2 + (z - r)^2}})}
%%%              {(R^2 + z^2 + \epsilon^2)^{5/2}} \, ,
%%%   \label{eq:plummer-smooth}
%%% \end{equation}
%%%%%%%%%%%%%%%%%%%%%%%%%%%%%%%%%%%%%%%%%%%%%%%%%%%%%%%%%%%%%%%%%%%%%%%%
where the second equality holds because the outer integral is taken
over the entire $z$ axis.}

\section{POWER-LAW PROFILES}

The first step is to examine the effect of Plummer smoothing on
power-law density \chng{profiles}, $\rho_n(r) = \rho_\mathrm{a} (a /
r)^n$, where $0 < n < 3$.
These profiles are not realistic, since the total mass diverges as $r
\to \infty$.
However, results obtained for power-law profiles help \chng{interpret}
the effects of smoothing on more realistic models.

\subsection{The case $\rho \propto r^{-1}$}

\begin{figure}
  \begin{minipage}{0.45\columnwidth}
    \begin{center}
    \includegraphics[clip=true,width=0.6\columnwidth]{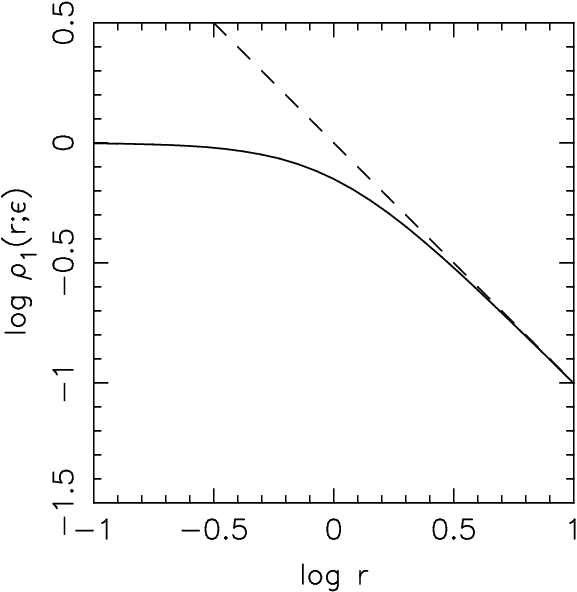}
    \caption{Effect of Plummer smoothing on a $\rho \propto r^{-1}$
    profile.
    Dashed line is the underlying density profile; solid curve is the
    result of smoothing with $\epsilon = 1$.
    The smoothed profile is always less than the underlying power-law.
    \label{cusp1}}
    \end{center}
  \end{minipage}%
  \hbox to 0.10\columnwidth{}%
  \begin{minipage}{0.45\columnwidth}
    \begin{center}
    \includegraphics[clip=true,width=0.6\columnwidth]{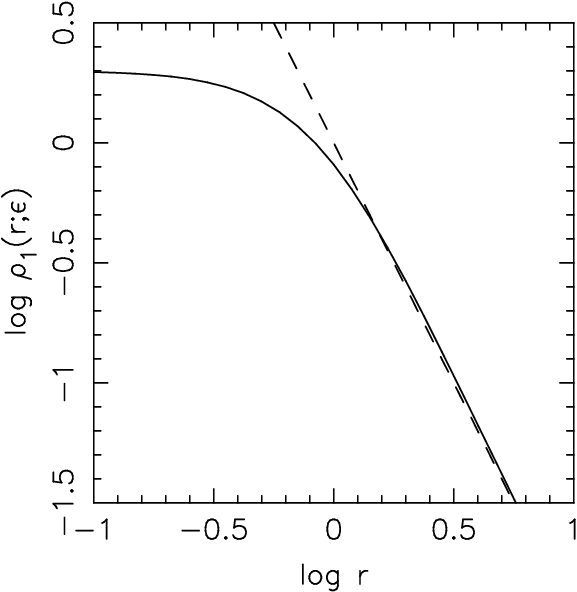}
    \caption{Effect of Plummer smoothing on a $\rho \propto r^{-2}$
    profile.
    Dashed line is the underlying density profile; solid curve is the
    result of smoothing with $\epsilon = 1$.
    The smoothed profile slightly exceeds the underlying
    power-law at large $r$.
    \label{cusp2}}
    \end{center}
  \end{minipage}%
\end{figure}

Let the density be $\rho_1(r) = \rho_\mathrm{a} (a / r)$.
The total mass enclosed within radius $r$ is $M_1(r) = 2 \pi
\rho_\mathrm{a} a r^2$.
In this case, the smoothed density profile can be calculated
analytically; the double integral is
\begin{equation}
  \int_{-\infty}^{\infty} dz
    \int_{0}^{\infty} dR \, R
      \frac{(R^2 + (z - r)^2)^{-1/2}}{(R^2 + z^2 + \epsilon^2)^{5/2}} =
        \frac{2}{3 \epsilon^2 \sqrt{\epsilon^2 + r^2}} \, .
\end{equation}
This yields a remarkably simple result for the smoothed
density, plotted in Fig.~\ref{cusp1},
\begin{equation}
  \rho_1(r;\epsilon) =
    \rho_\mathrm{a} \frac{a}{\sqrt{\epsilon^2 + r^2}} =
      \rho_1({\textstyle \sqrt{\epsilon^2 + r^2}}) = \rho_1(r_\epsilon) \, ,
\end{equation}
where $r_\epsilon \equiv \sqrt{\epsilon^2 + r^2}$.
The \chng{smoothed mass within radius $r$, hereafter called the
\textit{smoothed mass profile}}\footnote{\chng{This profile
can't be obtained by applying kernel smoothing directly to
$M(r)$; only density profiles can be smoothed.}}, is 
\begin{equation}
  M_1(r;\epsilon) =
    \int_{0}^{r} dx \, 4 \pi x^2 \, \rho_1(x;\epsilon) =
      2 \pi \rho_\mathrm{a} a
        \left( r r_\epsilon - \epsilon^2\sinh^{-1}(r/\epsilon) \right)
        \, .
\end{equation}

\subsection{The case $\rho \propto r^{-2}$}

Let the density be $\rho_2(r) = \rho_\mathrm{a} (a / r)^2$.
The total mass enclosed within radius $r$ is $M_2(r) = 4 \pi
\rho_\mathrm{a} a^2 r$.
The integral over \chng{$R$} can be evaluated, but the result is not
particularly informative and the remaining integral must be done
numerically.
Fig.~\ref{cusp2} presents the results.
For $\log(r) \ga 0.2$, the smoothed density exceeds the
underlying power-law profile.
This occurs because smoothing, \chng{in effect}, spreads mass from $r
\la \epsilon$ to larger radii, and with the underlying profile
dropping away so steeply this \chng{redistributed} mass makes a
relatively large contribution to $\rho_2(r;\epsilon)$.
Note that as $r \to 0$, the smoothed density $\rho_2(r;\epsilon) \to 2
\rho_2(\epsilon)$.

\subsection{Central density}

\begin{figure}
  \begin{center}
    \includegraphics[clip=true,width=0.45\columnwidth]{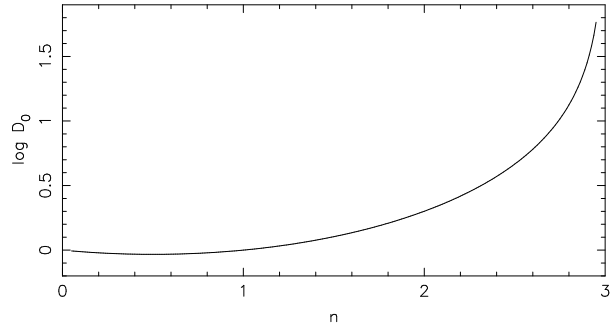}
    \caption{Density ratio $D_0 = \rho_n(0;\epsilon) /
    \rho_n(\epsilon)$ plotted as a function of $n$.
    Limiting values are $D_0 = 1$ as $n \to 0$ and $D_0 = \infty$ as
    $n \to 3$.
    \label{Dzero}}
  \end{center}
\end{figure}

It appears impossible to calculate the \chng{smoothed} density profile
for arbitrary $n$ without resorting to numerical methods, but the
central density is another matter.
Setting $r = 0$, the smoothed density is
\begin{equation}
  \rho_n(0;\epsilon) = 
    3 \epsilon^2 \int_{0}^{\infty} dx \,
                   \frac{x^2\rho_n(x)}{(x^2+\epsilon^2)^{5/2}} =
      \frac{n}{\sqrt{\pi}} \Gamma({\textstyle \frac{3}{2} - \frac{n}{2}}) 
        \Gamma({\textstyle \frac{n}{2}}) \rho_n(\epsilon) \, .
  \label{eq:central-rho-cusp}
\end{equation}
The central density ratio $D_0(n) = \rho_n(0;\epsilon) /
\rho_n(\epsilon)$ is plotted as a function of $n$ in Fig.~\ref{Dzero}.
For $n = 1$ and $2$, the ratio $D_0 = 1$ and $2$, respectively, in
accord with the results above, while as $n \to 3$ the central density
diverges.

The \chng{smoothed} central density for an arbitrary power-law is
useful in devising an approximate expression for the smoothed density
\chng{profile} (Appendix~A.1).
In addition, the central density \chng{is related to} the shortest
dynamical time-scale present in an $N$-body simulation, which may in
turn be used to estimate a maximum permissible value for the time-step
(\S~4.3.1).

\section{ASTROPHYSICAL MODELS}

\subsection{Hernquist and NFW models}

\begin{figure}
  \begin{minipage}{0.45\columnwidth}
    \begin{center}
      \includegraphics[clip=true,width=\columnwidth]{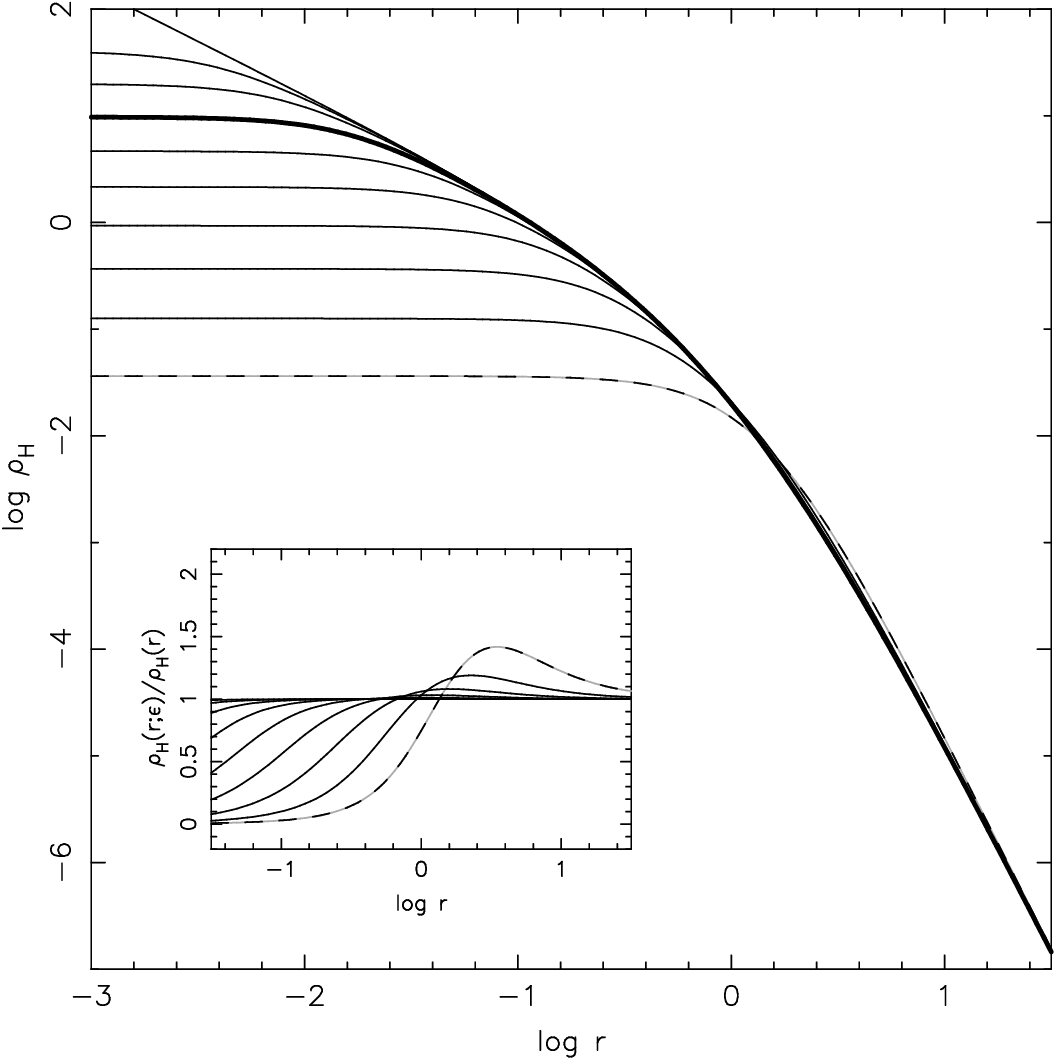}
      \caption{Effect of Plummer smoothing on Hernquist profile.
      Top curve shows the density profile of a Hernquist model with
      scale radius $a = 1$ and mass $M = 1$.
      Lower curves show profiles smoothed with $\epsilon = 1/256$,
      $1/128$, \dots, $1$ (from top to bottom); heavy curve is
      $\epsilon = 1/64$, \chng{dashed curve is $\epsilon = 1$.
      Inset shows ratio
      $\rho_\mathrm{H}(r;\epsilon)/\rho_\mathrm{H}(r)$}. 
      \label{rhoH}}
    \end{center}
  \end{minipage}%
  \hbox to 0.10\columnwidth{}%
  \begin{minipage}{0.45\columnwidth}
    \begin{center}
      \includegraphics[clip=true,width=\columnwidth]{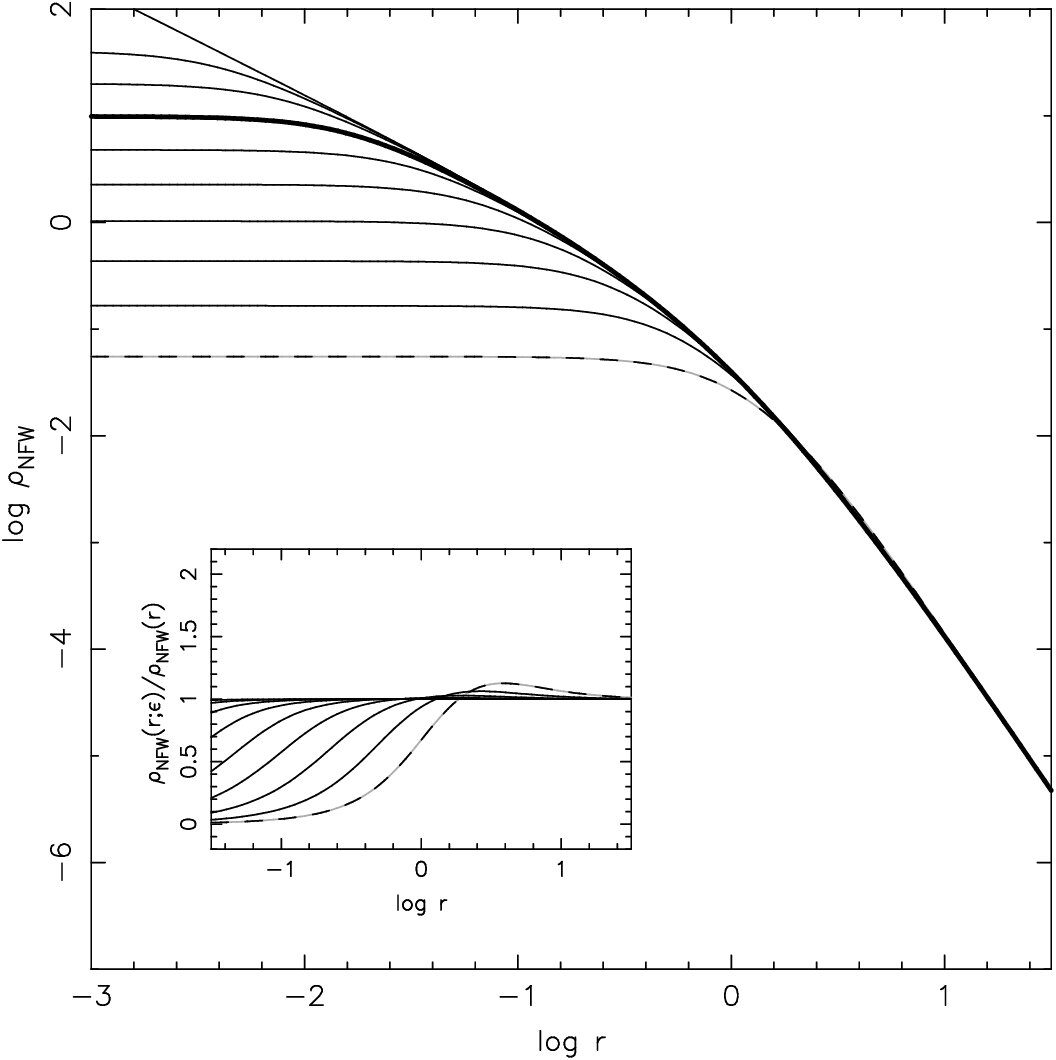}
      \caption{Effect of Plummer smoothing on NFW profile.
      Top curve shows the density profile of a NFW model with scale
      radius $a = 1$ and density $\rho_{0} = 1 / (2 \pi)$.
      Lower curves show profiles smoothed with $\epsilon = 1/256$,
      $1/128$, \dots, $1$ (from top to bottom); heavy curve is
      $\epsilon = 1/64$, \chng{dashed curve is $\epsilon = 1$.
      Inset shows ratio
      $\rho_\mathrm{NFW}(r;\epsilon)/\rho_\mathrm{NFW}(r)$}. 
      \label{rhoNFW}}
    \end{center}
  \end{minipage}%
\end{figure}

As noted above, both of these profiles have $\rho \propto r^{-1}$ as
$r \to 0$.
For this reason, they are treated in parallel.
The \citet{H90} model has density and mass profiles
\begin{equation}
  \rho_\mathrm{H}(r) = \frac{a M}{2 \pi r (a + r)^3} \, ,
  \qquad
  M_\mathrm{H}(r) = \frac{M r^2}{(r + a)^2} \, ,
  \label{eq:hernquist}
\end{equation}
where $a$ is the scale radius and $M$ is the total mass.
The \citet{NFW96} model has density and mass profiles
\begin{equation}
  \rho_\mathrm{NFW}(r) = \frac{a^3 \rho_{0}}{r (a + r)^2} \, ,
  \qquad
  M_\mathrm{NFW}(r) =
    4 \pi \rho_{0} a^3
      \left( \log\left(\frac{a + r}{a}\right) -\frac{r}{a + r} \right) \, ,
  \label{eq:NFW}
\end{equation}
where $a$ is again the scale radius and $\rho_{0}$ is a characteristic
density.
The double integrals required to evaluate the smoothed versions of
these profiles appear intractable analytically\footnote{\chng{Smoothed
\textit{central} densities for these and other profiles can be
expressed in terms of special functions.}} but can readily be
calculated numerically.
Figs.~\ref{rhoH} and~\ref{rhoNFW} present results for a range of
$\epsilon$ values between $a$ and $a/256$.
For comparison, both models are scaled to have the same
\chng{underlying density profile} at \chng{$r \ll a$}.

The \chng{smoothed profiles} shown in Figs.~\ref{rhoH}
and~\ref{rhoNFW} are, for the most part, easily understood in terms of
the results obtained for power-laws.
For radii $r < \epsilon$, smoothing transforms central cusps into
constant-density cores, just as in Fig.~\ref{cusp1}.
If the \chng{softening} length $\epsilon$ is much less than the scale
length $a$, the smoothed density $\rho(r;\epsilon)$ within $r \ll a$
is almost independent of the underlying profile at radii $r > a$.
Consequently, the smoothed central density $\rho(0;\epsilon) \simeq
\rho(\epsilon)$, echoing the result obtained for the power-law $n =
1$.
In addition, the actual curves in Figs.~\ref{rhoH} and~\ref{rhoNFW}
are shifted versions of the curves in Fig.~\ref{cusp1}; this
observation motivates simple approximations to
$\rho_\mathrm{H}(r;\epsilon)$ and $\rho_\mathrm{NFW}(r;\epsilon)$
described in Appendix~A.2.

On the other hand, if $\epsilon$ is comparable to $a$, the
quantitative agreement between these profiles and the smoothed $n = 1$
profile breaks down; the smoothed density at small $r$ has a
non-negligible contribution from the underlying profile beyond the
scale radius $a$.
As an example, for $\epsilon = 1$ the central density of the smoothed
NFW profile \chng{is higher than the central density} of the smoothed
Hernquist profile, because the former receives a larger contribution
from \chng{mass} beyond the scale radius.

A somewhat more subtle \chng{result, shown in the insets}, is that
\chng{heavily} smoothed profiles \textit{exceed} the underlying
profiles at radii $r \ga a$.
This is basically the same effect found with the $n = 2$ power-law
profile (\S~2.2); with the underlying density dropping rapidly as a
function of $r$, the mass spread outward from smaller radii more than
makes up for the mass spread to still larger radii.
This effect is more evident for the Hernquist profile than for the NFW
profile because the former falls off more steeply for $r \ga a$.

\subsection{Jaffe model}

\begin{figure}
  \begin{minipage}[t]{0.45\columnwidth}
    \begin{center}
      \includegraphics[clip=true,width=\columnwidth]{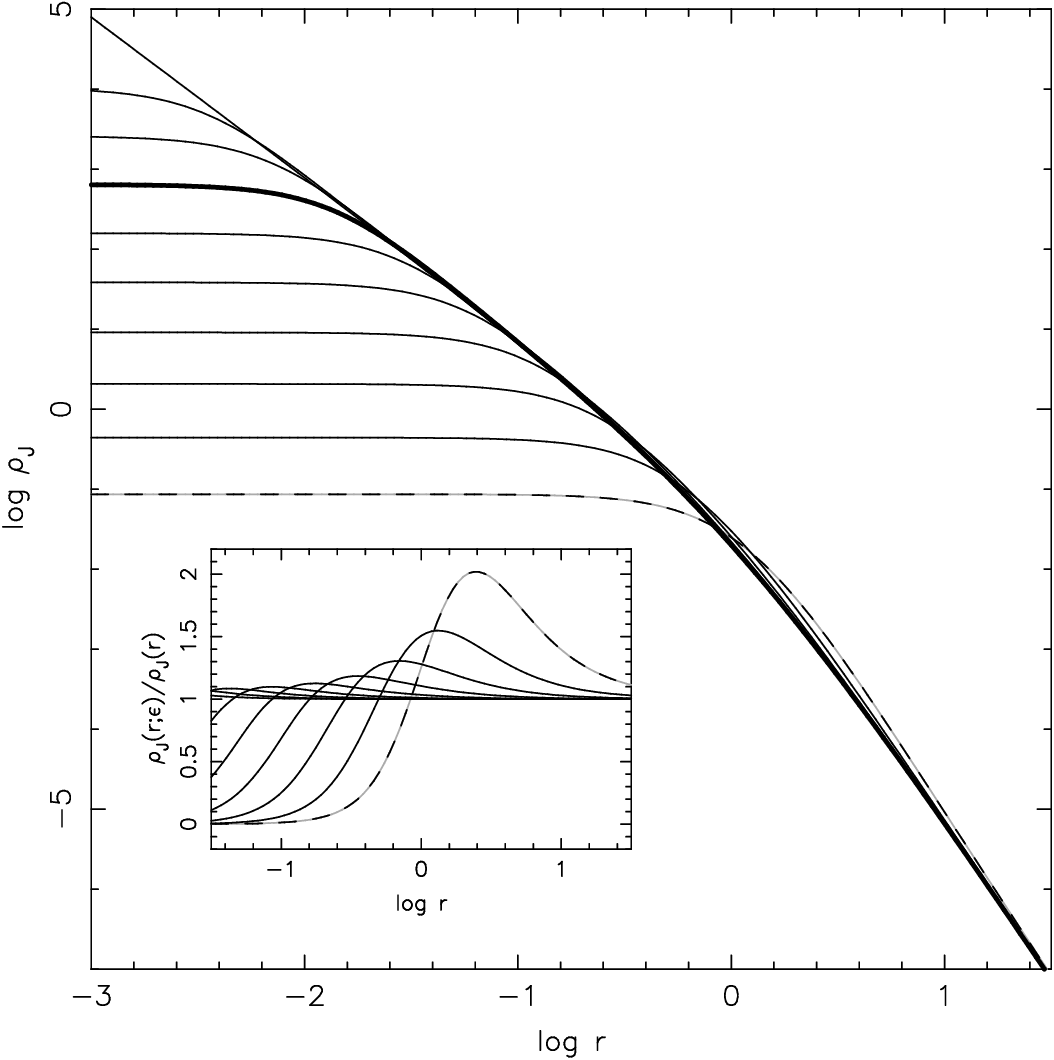}
      \caption{Effect of Plummer smoothing on Jaffe profile.
      Top curve shows the density profile of a Jaffe model with scale
      radius $a = 1$ and mass $M = 1$.
      Lower curves show profiles smoothed with $\epsilon = 1/256$,
      $1/128$, \dots, $1$ (from top to bottom); heavy curve is
      $\epsilon = 1/64$, \chng{dashed curve is $\epsilon = 1$.
      Inset shows ratio
      $\rho_\mathrm{J}(r;\epsilon)/\rho_\mathrm{J}(r)$.}
      \label{rhoJ}}
    \end{center}
  \end{minipage}%
  \hbox to 0.10\columnwidth{}%
  \begin{minipage}[t]{0.45\columnwidth}
    \begin{center}
      \includegraphics[clip=true,width=\columnwidth]{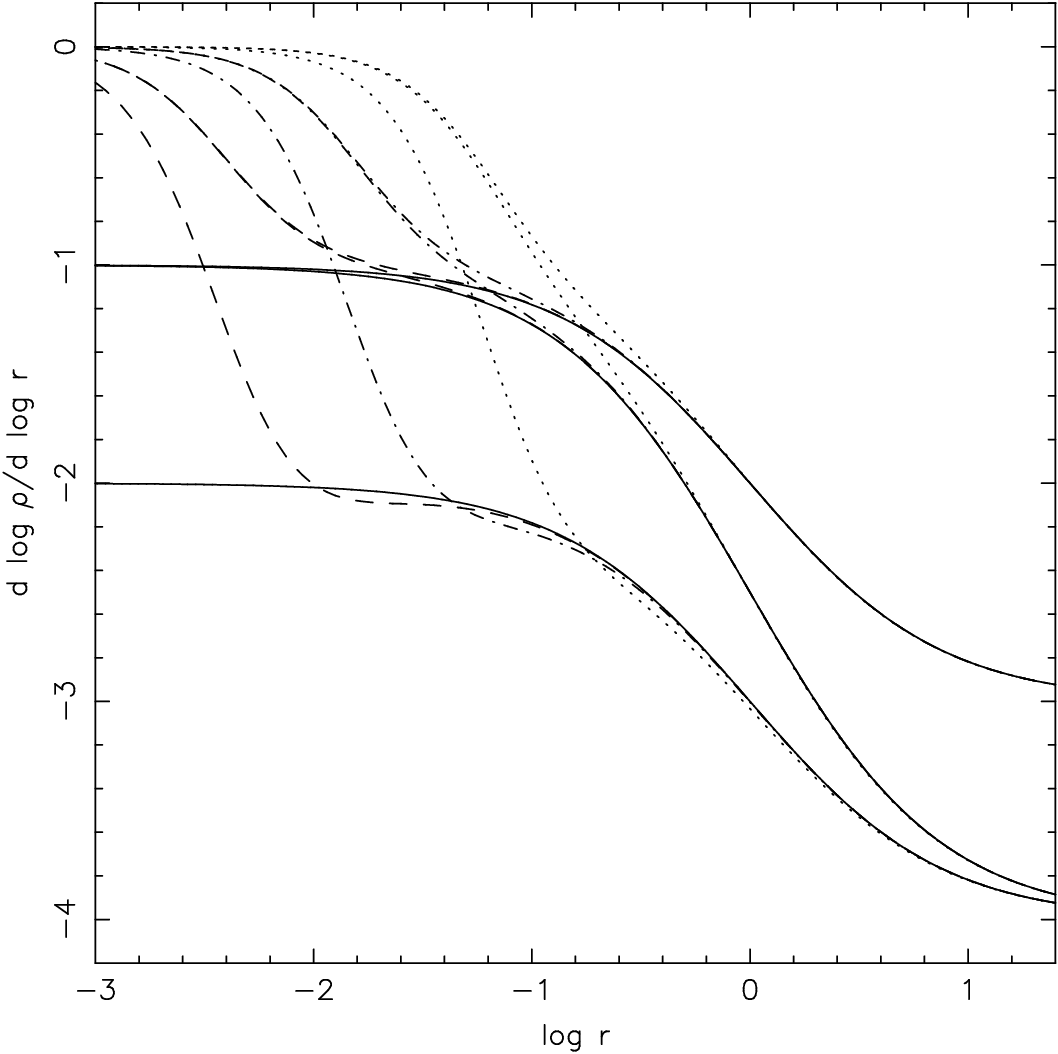}
      \caption{Logarithmic slopes of profiles from Figs.~\ref{rhoH},
      \ref{rhoNFW}, and~\ref{rhoJ}.
      Solid lines are underlying profiles; from bottom to top, they
      represent Jaffe, Hernquist, and NFW models, respectively.
      Dashed, dot-dashed, and dotted lines give results for $\epsilon
      = a/256$, $a/64$, and $a/16$, respectively.
      \label{slope}}
    \end{center}
  \end{minipage}%
\end{figure}

The \citet{J83} model has density and mass profiles
\begin{equation}
  \rho_\mathrm{J}(r) = \frac{a M}{4 \pi r^2 (a + r)^2} \, ,
  \qquad
  M_\mathrm{J}(r) = \frac{M r}{(r + a)} \, ,
  \label{eq:jaffe}
\end{equation}
where $a$ is the scale radius and $M$ is the total mass.
The double integrals required to evaluate the smoothed version of this
profile appear intractable analytically but can readily be calculated
numerically.
Fig.~\ref{rhoJ} present results for a range of $\epsilon$ values
between $a$ and $a/256$.

Again, much of the behavior shown in this plot can be understood by
reference to the results for the $n = 2$ power-law.
In particular, for smoothing lengths $\epsilon \ll a$, the central
density is $\rho_\mathrm{J}(0;\epsilon) \simeq 2
\rho_\mathrm{J}(\epsilon)$, and the curves in Fig.~\ref{rhoJ} are
shifted versions of the one in Fig.~\ref{cusp2}.
\chng{As the inset shows}, for larger values of $\epsilon$ the
smoothed profiles quite noticeably exceed the underlying profile; the
effect is stronger here than it is for a Hernquist model because the
Jaffe model has more mass \chng{within $r \le a$ to redistribute}.

\subsection{How much softening is too much?}

Figs.~\ref{rhoH}, \ref{rhoNFW}, and~\ref{rhoJ} have interesting
implications for $N$-body experiments.
One might expect the smoothed \chng{profiles} to resolve the inner
power-laws of the underlying models as long as the softening length
$\epsilon$ is somewhat less than the scale radius $a$, but \chng{that}
is not \chng{what these figures show}.
Profiles smoothed with $\epsilon \ga a/16$ are essentially
constant-density cores attached to power-law outer profiles; the
density within the core depends on $\epsilon$, but no inner cusp per
se can be seen.
For $\epsilon \la a/64$, on the other hand, the smoothed profiles do
appear to trace the inner power-laws over some finite range of radii,
before flattening out at smaller $r$.
Only for $\epsilon \la a/256$ can the inner cusps be followed for at
least a decade in radius.

Fig.~\ref{slope} helps explain this result.
The underlying Jaffe, Hernquist, and NFW profiles all roll over
gradually from their inner to outer power-law slopes between radii
$0.1 a \la r \la 10 a$.
Thus a resolution somewhat better than $0.1 a$ \chng{is required} to
see the inner cusps of these models.
In practice, this implies \chng{the} softening parameter $\epsilon$
\chng{must be} several times smaller than $0.1 a$.

\section{TESTS AND APPLICATIONS}

Since the formalism developed above is exact, numerical tests of a
relation like (\ref{eq:plummer-smooth}) for the smoothed density
$\rho(r;\epsilon)$ may seem superfluous.
In practice, such tests can be illuminating -- as benchmarks of
$N$-body technique.
In what follows, the smoothing formalism will be applied to actual
$N$-body calculations, to check $N$-body methodology and to
demonstrate that the formalism has real applications.

Putting this plan into operation requires some care.
To begin with, an $N$-body realization of a standard Hernquist or
Jaffe profile spans a huge range of radii.
Typically, the innermost particle has radius $r_\mathrm{in} \sim a
N^{-1/2}$ or $a N^{-1}$ for a Hernquist or Jaffe profile,
respectively, while for either profile, the outermost particle has
radius $r_\mathrm{out} \sim a N$.
A dynamic range of $r_\mathrm{out}/r_\mathrm{in} \sim N^{3/2}$
or~$N^2$ can be awkward to handle numerically; even gridless tree
codes may not accommodate such enormous ranges gracefully.
\chng{One simple option is to} truncate the particle distribution at
some fairly large radius, but \chng{it's preferable to smoothly} taper
the density profile:
\begin{equation}
  \rho(r) \to \rho_\mathrm{t}(r) =
    \left\{
      \begin{array}{ll}
	\!\! (1 + \mu) \, \rho(r) \, , &
	    r \le b \\ [0.125in]
	\!\! (1 + \mu) \, \rho_{*} \, (b / r)^2 \, e^{-r / r_{*}} \, , &
	    r > b \\
      \end{array}
    \right.
  \label{eq:tapered-models}
\end{equation}
where the taper radius $b \gg a$, the values of $r_{*}$ and $\rho_{*}$
are fixed by requiring that $\rho_\mathrm{t}(r)$ and its first
derivative are continuous at $r = b$, and the value of $\mu \ll 1$ is
chosen to preserve the total mass.
Let
\begin{equation}
  \beta = \frac{r}{\rho} \left . \frac{d \rho}{d r} \right|_{r = b}
\end{equation}
be the logarithmic slope of the density profile at $r = b$, \chng{and
$M(r)$ be the underlying mass profile}; then
\begin{equation}
  r_{*} = \frac{b}{-(2 + \beta)} \, ,
  \qquad
  \rho_{*} = \rho(b) e^{-(2 + \beta)} \, ,
  \qquad \mathrm{and} \qquad
  \mu = \frac{M(\infty)}{M(b) + 4 \pi b^2 r_{*} \rho(b)} - 1 \, .
\end{equation}

\begin{figure}
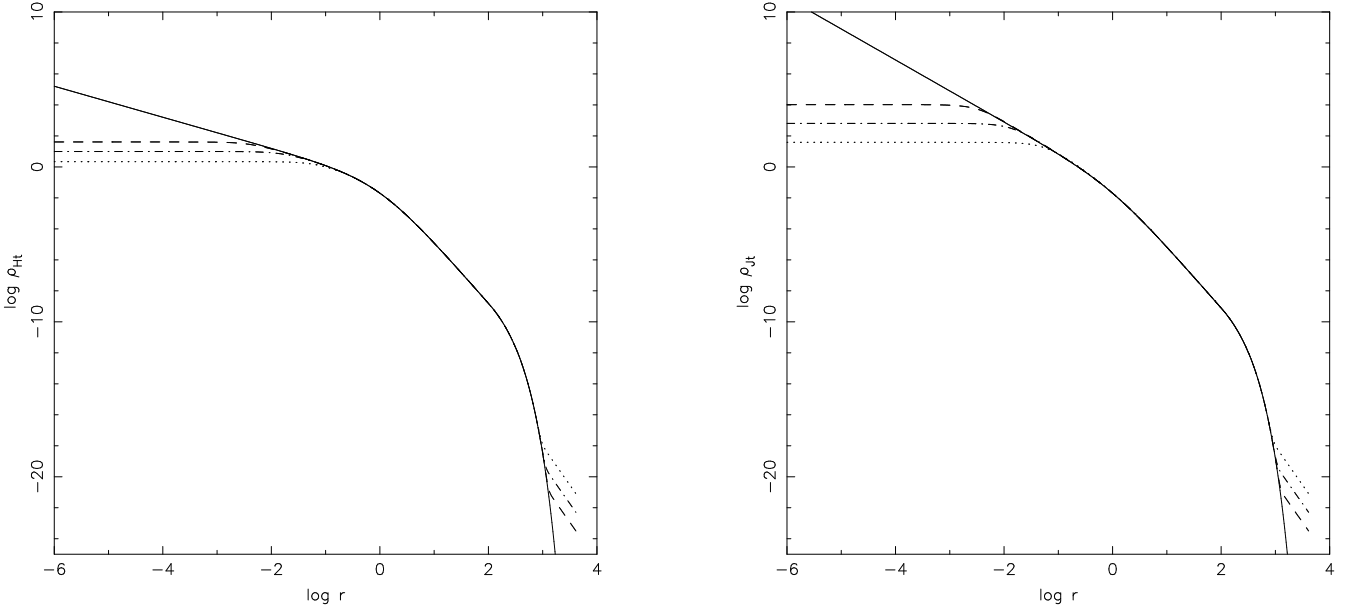

  \begin{minipage}{0.45\columnwidth}
    \includegraphics[clip=true,width=\columnwidth]{rhoHt.ps}
  \end{minipage}%
  \hbox to 0.10\columnwidth{}%
  \begin{minipage}{0.45\columnwidth}
    \includegraphics[clip=true,width=\columnwidth]{rhoJt.ps}
  \end{minipage}%
  \caption{Effect of Plummer smoothing on tapered Hernquist (left) and
  Jaffe (right) models, represented by the solid curves.
  Dashed, dot-dashed, and dotted curves show results for $\epsilon =
  1/256$, $1/64$, and $1/16$, respectively.
  \label{rhoHJt}}
\end{figure}

Fig.~\ref{rhoHJt} shows how Plummer smoothing modifies tapered
Hernquist and Jaffe profiles.
Both profiles have scale radius $a = 1$, taper radius $b = 100$, and
mass $M = 1$; these parameters will be used in all subsequent
calculations.
In each case, the underlying profile follows the standard curve out to
the taper radius $b$, and then rapidly falls away from the outer power
law.
At radii $r \le b$, the smoothed profiles match those shown in
Figs.~\ref{rhoH} and \ref{rhoJ}, apart from the factor of $(1 + \mu)$
used to preserve total mass.
At larger radii, the smoothed profiles initially track the underlying
tapered profiles, but then transition to asymptotic $\rho \propto
r^{-5}$ power law tails.
This occurs because the Plummer smoothing kernel
(\ref{eq:plummer-kernel}) falls off as $r^{-5}$ at large $r$; in
fact, these power laws match $\rho = 3 M \epsilon^2 / 4 \pi r^5$,
which is the large-$r$ approximation for a point mass $M$ smoothed
with a Plummer kernel.
The amount of mass in these $r^{-5}$ tails is negligible.

\subsection{Gravitational potentials}

In principle, it's straightforward to verify that the smoothed
profiles above generate potentials matching those obtained from $N$-body
calculations.
For a given density profile $\rho(r)$, construct a realization with
$N$ \chng{particles} at positions $\vect{r}_i$; a $N$-body force
calculation with softening $\epsilon$ yields the gravitational
potential $\Phi_i$ for each particle.
Conversely, given the smoothed \chng{density} profile
$\rho(r;\epsilon)$, compute the smoothed mass profile $M(r;\epsilon)$,
and use \chng{the result} to obtain the \textit{\chng{smoothed}
potential} $\Phi(r;\epsilon)$:
\begin{equation}
  \frac{d\Phi}{dr} = G \frac{M(r;\epsilon)}{r^2} \, ,
  \label{eq:potential-equation}
\end{equation}
with boundary condition $\Phi \to 0$ as $r \to \infty$.
For each particle, the $N$-body potential $\Phi_i$ may be compared
with the predicted value $\Phi(|\vect{r}_i|;\epsilon)$; apart from
$\sqrt{N}$ fluctuations, the two should agree.

A major complication is that $\sqrt{N}$ fluctuations in $N$-body
realizations imprint spatially coherent perturbations on the
gravitational field; potentials measured at adjacent positions are not
statistically independent.
For example, the \chng{softened} potential at the origin of an
$N$-body system is
\begin{equation}
  \Phi_0 = \sum_i \frac{G m_i}{(r_i^2 + \epsilon^2)^{1/2}} \, .
  \label{eq:$N$-body-central-potential}
\end{equation}
If the radii $r_i$ are independently chosen, this expression is a
Monte Carlo integral, which will deviate from $\Phi(0;\epsilon)$ by an
amount of order $O(N^{-1/2})$; moreover, everywhere within $r \la
\epsilon$ the potential will deviate upward or downward by roughly as
much as it does at $r = 0$.
One way around this is to average over many $N$-body realizations, but
this is tedious and expensive.
An easier solution is to sample the radial distribution uniformly.
Let $M(r)$ be the mass profile associated with the underlying density
$\rho(r)$.
\chng{Assign all particles equal masses, and determine the radius
$r_i$ of particle $i$} by solving $M(r_i) = (i - 0.5) M(\infty)/N$ for
$i = 1$ to $N$.
This eliminates radial fluctuations; the Monte-Carlo
integral for $\Phi_0$ is replaced with a panel integration uniformly
spaced in $M(r)$, and the central potential is obtained with
relatively high accuracy.

This trick does not suppress \textit{non-radial} fluctuations, so the
$N$-body potential evaluated at any \chng{point} $r > 0$ still differs
from the true $\Phi(r;\epsilon)$.
But a non-radial fluctuation which creates an overdensity at some
position $\vect{r}_\mathrm{over}$ must borrow mass from elsewhere on
the sphere $r = \chng{|\vect{r}_\mathrm{over}|}$; over-dense and
under-dense regions compensate each other when averaged over the
entire surface of the sphere.
The resulting potential fluctuations likewise average to zero over the
sphere, \chng{as one can show by using Gauss's theorem to evaluate the
average gradient of the potential and integrating inward from $r =
\infty$.}

Finally, a subtle bias arises if the particles used to generate the
potential are also used to probe the potential, since local
overdensities are sampled more heavily.
To avoid this, the potential can be measured at a set of points
$\vect{r}_k$ which are \textit{independent} of the particle positions
$\vect{r}_i$.
Then $\delta\Phi_k \equiv \Phi_k - \Phi(r_k;\epsilon)$ should display
some scatter, but average to zero when integrated over test points
$\vect{r}_k$ within a spherical shell.

\begin{figure}
  \begin{minipage}{0.45\columnwidth}
    \includegraphics[clip=true,width=\columnwidth]{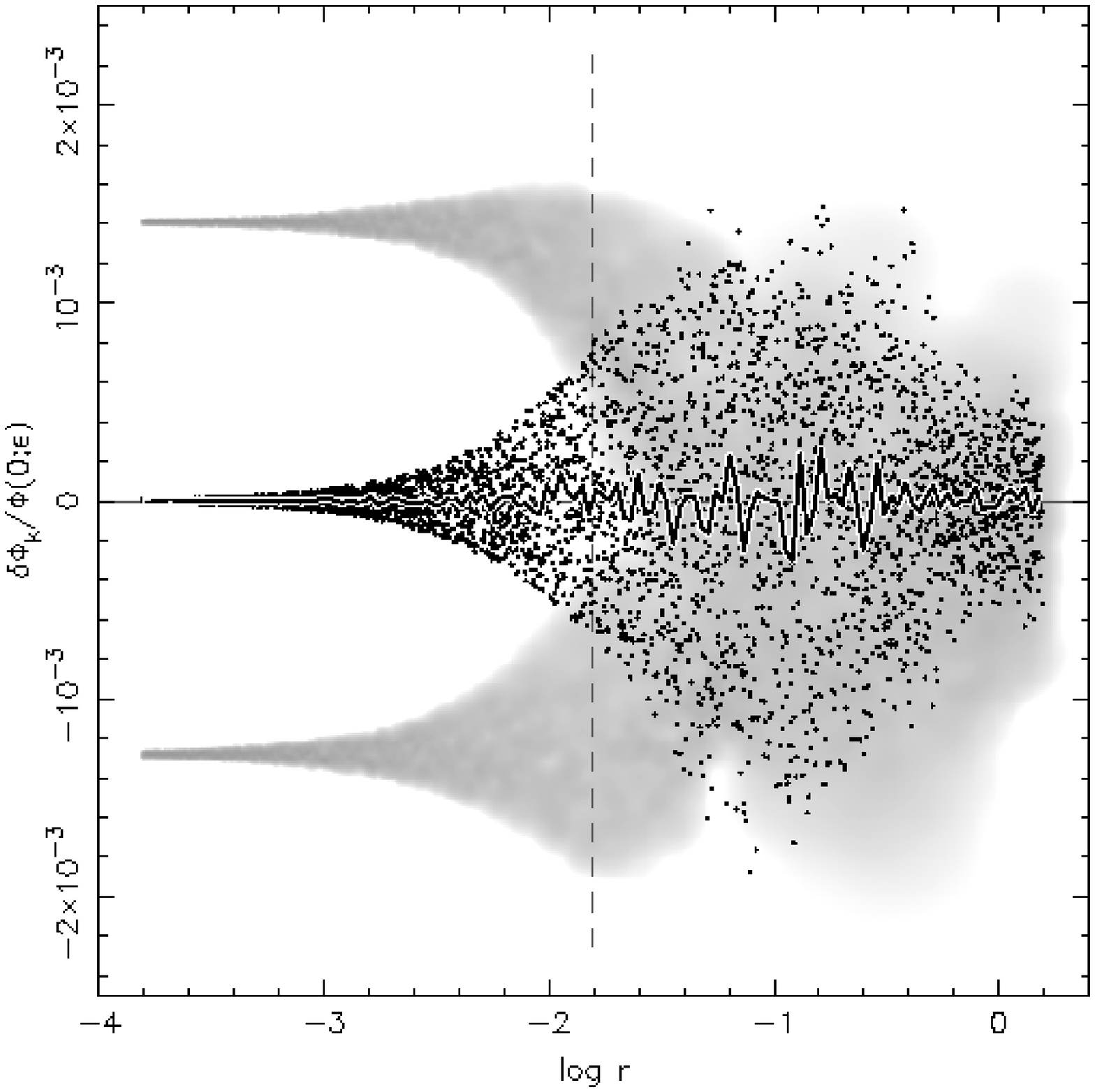}
  \end{minipage}%
  \hbox to 0.10\columnwidth{}%
  \begin{minipage}{0.45\columnwidth}
    \includegraphics[clip=true,width=\columnwidth]{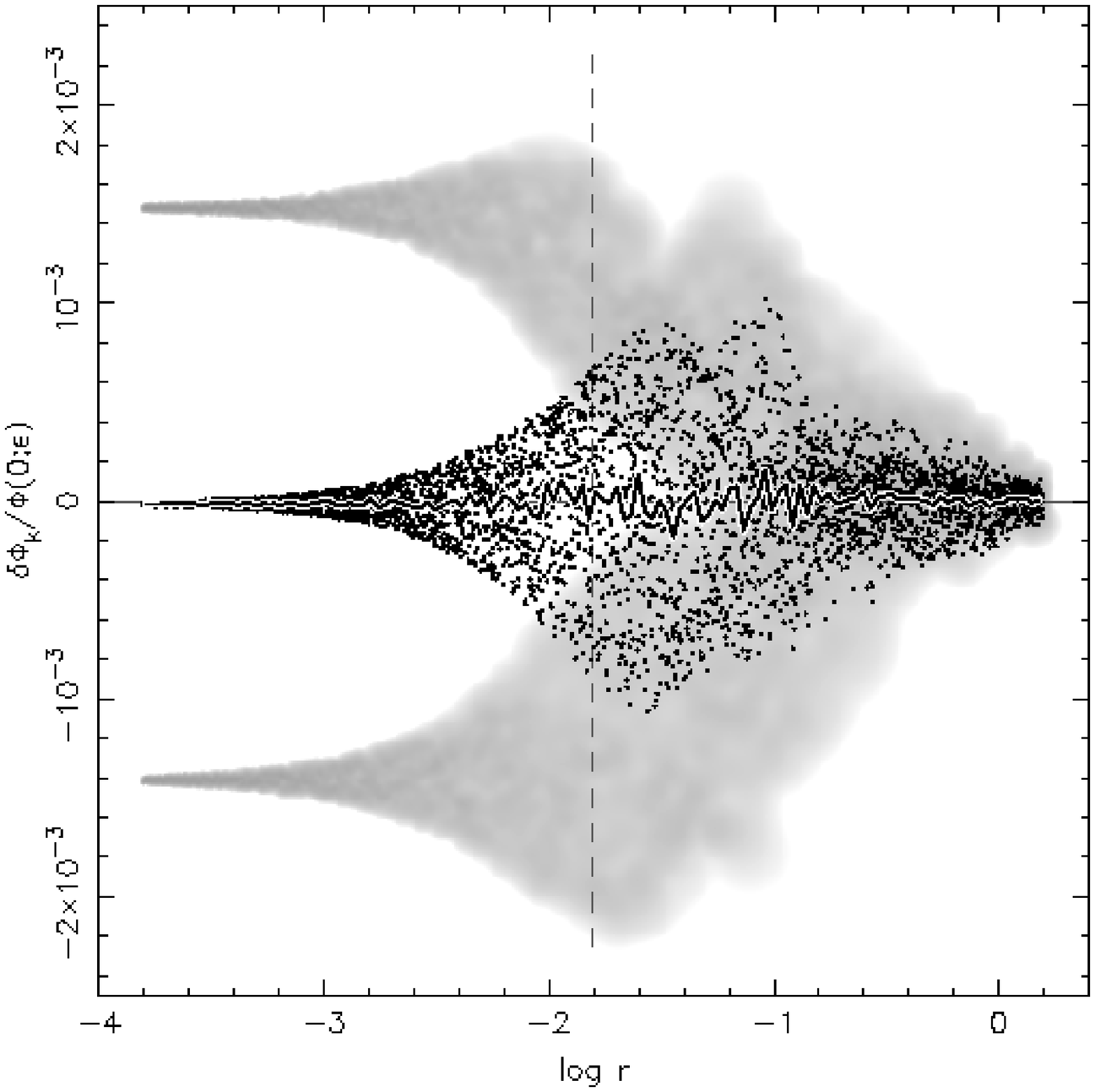}
  \end{minipage}%
  \caption{Difference $\delta\Phi_k$ between $N$-body and smoothed
  potentials for tapered Hernquist (left) and Jaffe (right) models,
  normalized by central potential $\Phi(0;\epsilon)$.
  \chng{Vertical dashed} lines show value of $\epsilon = 1/64$.
  Points are results for uniform realizations; \chng{jagged} curves
  shows averages for groups of 32 points.
  \chng{Grey-scale images show} typical results for \textit{random}
  realizations.
  Central potentials are $\Phi_\mathrm{H}(0;\epsilon) = -0.9809$ and
  $\Phi_\mathrm{J}(0;\epsilon) = -3.9009$.
  \label{deltaPhiHJ}}
\end{figure}

Fig.~\ref{deltaPhiHJ} shows results from direct-sum potential
calculations for tapered Hernquist and Jaffe models, using units with
$G = 1$.
In each case, the underlying density profile was represented with $N =
2^{20} = 1048576$ equal-mass \chng{particles}, and the potential was
measured at $4096$ points uniformly distributed in $\log(r)$ between
$0.01 \epsilon$ and $100 \epsilon$.
The points show results for uniform radial sampling.
While non-radial fluctuations create scatter in $\Phi_k$, the
distribution is fairly symmetric about the line $\delta\Phi = 0$.
The jagged curves are obtained by averaging $\delta\Phi_k$ over radial
bins each containing $32$ points.
These averages \chng{fall} near zero, demonstrating very good
agreement between the $N$-body results and the potentials calculated
from the smoothed density profiles.

For comparison, the \chng{grey-scale images} in Fig.~\ref{deltaPhiHJ}
\chng{display representative} results for \textit{random} realizations
of \chng{each} density profile.
In these realizations, the radius of particle $i$ is computed by
solving $M(r_i) = X_i M(\infty)$, where $X_i$ is a random
number\footnote{A good random number generator is essential.  The Unix
generator, \texttt{random()}, appears to be slightly non-uniform;
replacing $X_i$ with $1 - X_i$ yields systematically different
$\Phi_0$ values.  The results shown here use the Tausworthe generator
\texttt{taus2} \citep{G+09}.} uniformly distributed between $0$ and
$1$.
To \chng{examine} the range of outcomes, $1000$ random realizations of
each model were generated and ranked by central potential $\Phi_0$;
since particle radii are chosen independently, the central limit
theorem implies that $\Phi_0$ has a normal distribution.
\chng{Shown} here are the $25^\mathrm{th}$ percentile and
$75^\mathrm{th}$ percentile members of these ensembles; half of all
random realizations lie between the two examples presented in each
figure.
Note that these examples deviate from the true potential by fractional
amounts of $\sim N^{-1/2}$.
Obviously, it's impossible to detect discrepancies between $\Phi_k$
and $\Phi(r_k;\epsilon)$ of less than one part in $10^3$ using random
realizations with $N \sim 10^6$.

It's instructive, not to mention disconcerting, to try reproducing
Fig.~\ref{deltaPhiHJ} using a tree code \citep{BH86} instead of direct
summation.
Tree codes employ approximations which become less accurate for
$\epsilon > 0$ \citep{H87, WC06}; these systematically bias computed
potentials and accelerations (see Appendix~B).
For example, the code which will shortly be used for dynamical tests,
run with an opening angle $\theta = 0.8$, yields central potentials
which are too deep by a few parts in $10^3$, depending on the system
being modeled.
This systematic error cannot be `swept under the carpet' when
comparing computed and predicted potentials at the level of precision
attempted here.

\subsection{Distribution functions}

Constructing equilibrium configurations is an important element of
many $N$-body experiments.
Approximate equilibria may be generated by a variety of ad hoc
methods, but the construction of a true equilibrium $N$-body model
amounts to drawing particle positions and velocities from an
equilibrium distribution function $f = f(\vect{r},\vect{v})$.
However, a configuration based on a distribution function (DF) derived
without \chng{allowing for} softening will \textit{not} be in
equilibrium if it is simulated with softening.

Assume the model to be constructed is spherical and isotropic.
Broadly speaking, there are two options: (a) adopt a DF $f = f(E)$
which depends on the energy $E$, and solve Poisson's equation for the
gravitational potential, or (b) adopt a mass model $\rho = \rho(r)$,
and use Eddington's (\citeyear{E16}) formula to solve for the DF.
If softening is taken into account, option (a) becomes \chng{somewhat
awkward, since} the source term for Poisson's equation
(\ref{eq:smooth-poisson}) is non-local\footnote{\chng{\citet{DS00}
describe an iterative scheme using softened $N$-body potentials which
implements option (a).}}.
\chng{On the other hand}, option (b) is relatively straightforward
\citep[e.g.,][]{KMM04}.

Starting with a desired density profile $\rho(r)$, the first step is
to compute the smoothed density and mass profiles $\rho(r;\epsilon)$
and $M(r;\epsilon)$, respectively.
Since $M(r;\epsilon) \ge 0$ everywhere, equation
(\ref{eq:potential-equation}) guarantees that the \chng{smoothed}
potential $\Phi(r;\epsilon)$ is a monotonically increasing function of
$r$.
It is therefore possible to express the underlying density profile
$\rho(r)$ as a function of $\Phi(r;\epsilon)$, and compute the DF:
\begin{equation}
  f(E;\epsilon) = \frac{1}{\sqrt{8} \pi^{2}} \, \frac{d}{dE}
    \int_{E}^{0} d\Phi \, (\Phi - E)^{-1/2} \, \frac{d\rho}{d\Phi} \, .
  \label{eq:eddington-formula}
\end{equation}
Note that \chng{in $d\rho/d\Phi$, the quantity} $\rho = \rho(r)$ is
the \textit{underlying} density, while $\Phi = \Phi(r;\epsilon)$
\chng{is the \textit{smoothed} potential}, related by Poisson's
equation to the smoothed density $\rho(r;\epsilon)$.
In effect, the \chng{smoothed} potential $\Phi(r;\epsilon)$ is taken
as a given, and (\ref{eq:eddington-formula}) is used to find what will
hereafter be called the \chng{\textit{smoothed distribution
function}} $f(E;\epsilon)$; \chng{with this DF}, the underlying
profile $\rho(r)$ \chng{is} in equilibrium in \chng{the} adopted
potential \citep[e.g.,][]{MD07,BH09}.
\chng{Conversely, setting $\epsilon = 0$ yields the
\textit{self-consistent distribution function} $f(E)$ which describes
a self-gravitating model with the underlying profile $\rho(r)$.}

\begin{figure}
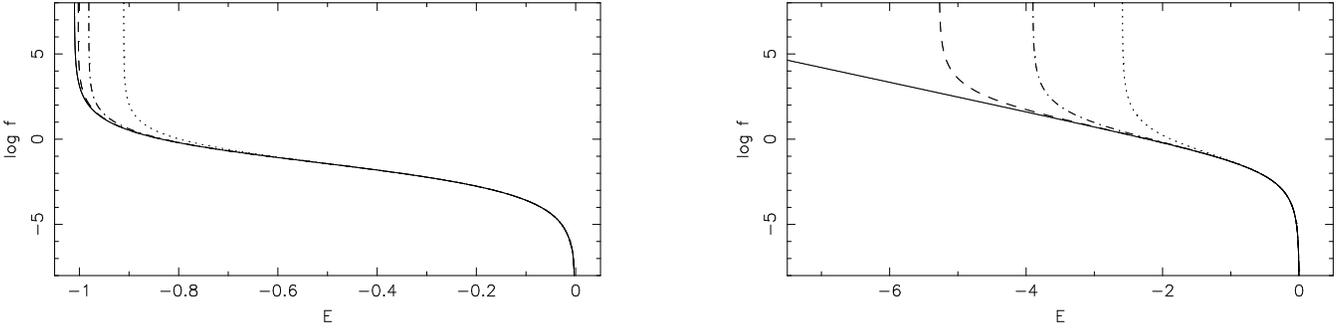

  \begin{minipage}{0.45\columnwidth}
    \includegraphics[clip=true,width=\columnwidth]{dfH.ps}
  \end{minipage}%
  \hbox to 0.10\columnwidth{}%
  \begin{minipage}{0.45\columnwidth}
    \includegraphics[clip=true,width=\columnwidth]{dfJ.ps}
  \end{minipage}%
  \caption{Distribution functions for tapered Hernquist (left) and
  Jaffe (right) models.
  Solid curve: $f(E)$; dashed, dot-dashed, and dotted curves:
  $f(E;\epsilon)$ for $\epsilon = 1/256$, $1/64$, and $1/16$,
  respectively.
  \label{dfHJ}}
\end{figure}

Fig.~\ref{dfHJ} presents DFs for tapered Hernquist and Jaffe models.
In each case the solid line shows \chng{the self-consistent DF}; these
match the published DFs \citep{J83, H90} over almost the entire energy
range, deviating only for $E \ga -G M / b = -0.01$ where tapering sets
in.

\chng{Smoothed DFs for \chng{Jaffe} models appear very different from
their self-consistent counterpart.}
The \chng{latter} has a logarithmic, infinitely deep
potential well, which effectively confines material with
\chng{constant} velocity dispersion in a $\rho \propto r^{-2}$ cusp.
The characteristic phase-space density $f \sim \rho
\sigma^{-3} \propto r^{-2}$ diverges as $r \to 0$ (ie, as $E \to
-\infty$), but only because $\rho$ does.
With $\epsilon > 0$ the potential well is harmonic at small $r$, and
can't confine a $\rho \propto r^{-2}$ cusp unless the local velocity
dispersion scales as $\sigma \propto r$; thus the phase-space density
now diverges as $f \propto r^{-5}$.
Moreover, the domain of $f(E;\epsilon)$ is limited
to to $E_0 \le E \le 0$, where $E_0 \equiv \Phi(0;\epsilon)$.
Thus, instead of growing exponentially as a function of $-E$, the
\chng{smoothed} DF abruptly diverges at some finite energy.

By comparison, \chng{smoothed} DFs for \chng{Hernquist} models \chng{look
similar to the self-consistent DF}.
The latter has a potential well of finite depth, and the smoothed
profiles generate wells which are only slightly shallower.
As the left panel of Fig.~\ref{dfHJ} shows, all the DFs asymptote to
$\infty$ as $E \to E_0$.
However, the run of velocity dispersion with $r$ \chng{is different};
the self-consistent model has $\sigma \propto r^{1/2}$, implying $f
\propto \rho \sigma^{-3} \propto r^{-5/2}$.
In contrast, the smoothed models have $\sigma \propto r$, implying $f
\propto r^{-4}$.

One consequence is that the way in which $f \to \infty$ as $E \to E_0$
\chng{is different in} the smoothed and \chng{self-consistent}
Hernquist models. 
The \chng{self-consistent} model has a linear potential as small $r$,
and thus $f \propto r^{-5/2} \propto (E - E_0)^{-5/2}$.
By comparison, the models based on smoothed potentials have harmonic
cores, and as a result, $f \propto r^{-4} \propto (E - E_0)^{-2}$.
(This difference is not apparent in Fig.~\ref{dfHJ} but becomes
obvious when $\log (E - E_0)$ is plotted against $\log f$.)
In this respect, the use of a smoothed potential \chng{effects} a
non-trivial change on Hernquist models: $f$ is a different power-law
of $(E - E_0)$.
Coincidentally, the smoothed Jaffe models have $f \propto r^{-5}
\propto (E - E_0)^{-5/2}$, just like the \chng{self-consistent}
Hernquist model.

\subsection{Dynamical tests}

$N$-body simulations are useful to show that the distribution
functions just constructed are actually in dynamical equilibrium with
their smoothed potentials.
For each model and $\epsilon$ value, two ensembles of three random
realizations were run.
In one ensemble, the initial conditions were generated using the
\chng{self-consistent} DF $f(E)$.
The other ensemble used initial conditions generated from the
\chng{smoothed} DF $f(E;\epsilon)$, which allows for the effects of
softening.

Each realization contained $N = 2^{18} = 262144$ equal-mass particles.
Initial particle radii $r_i$ were selected randomly \chng{by solving
$M(r_i) = X_i M(\infty)$} as described above.
\chng{Initial particle speeds $v_i$ were selected randomly by
rejection sampling \citep{vN51} from the distributions $g(v;r_i) = v^2
f(\frac{1}{2}v^2 + \Phi(r_i))$ or $g(v;r_i) = v^2 f(\frac{1}{2}v^2 +
\Phi(r_i;\epsilon);\epsilon)$, where the former assumes the
self-consistent DF, and the latter a smoothed DF.
Position and velocity vectors for particle $i$ are obtained by
multiplying $r_i$ and $v_i$ by independent unit vectors drawn from an
isotropic distribution.
In effect, this procedure treats the 6-D distribution function
$f(\vect{r},\vect{v})$ as a probability density, and selects each
particle's coordinates independently.}

Simulations were run using a hierarchical $N$-body code\footnote{See
\texttt{http://www.ifa.hawaii.edu/faculty/barnes/treecode/treeguide.html}
for a description.}.
An opening angle of $\theta = 0.8$, together with quadrupole moments,
provided forces with median errors $\delta\vect{a}/|\vect{a}| \la
0.0006$.
Particles within \chng{$r \sim 10 \epsilon$ have} much larger force
errors, although these seem to have \chng{limited} effect in practice
(Appendix~B).
Trajectories were integrated using a time-centered leap-frog, with the
same time-step $\Delta t = 1/1024$ for all particles (see \S~4.3.1).
\chng{All simulations were run to $t = 16$, which is more than
sufficient to test initial equilibrium.}

\begin{figure}
  \begin{center}
    \includegraphics[clip=true,width=\columnwidth,angle=0]{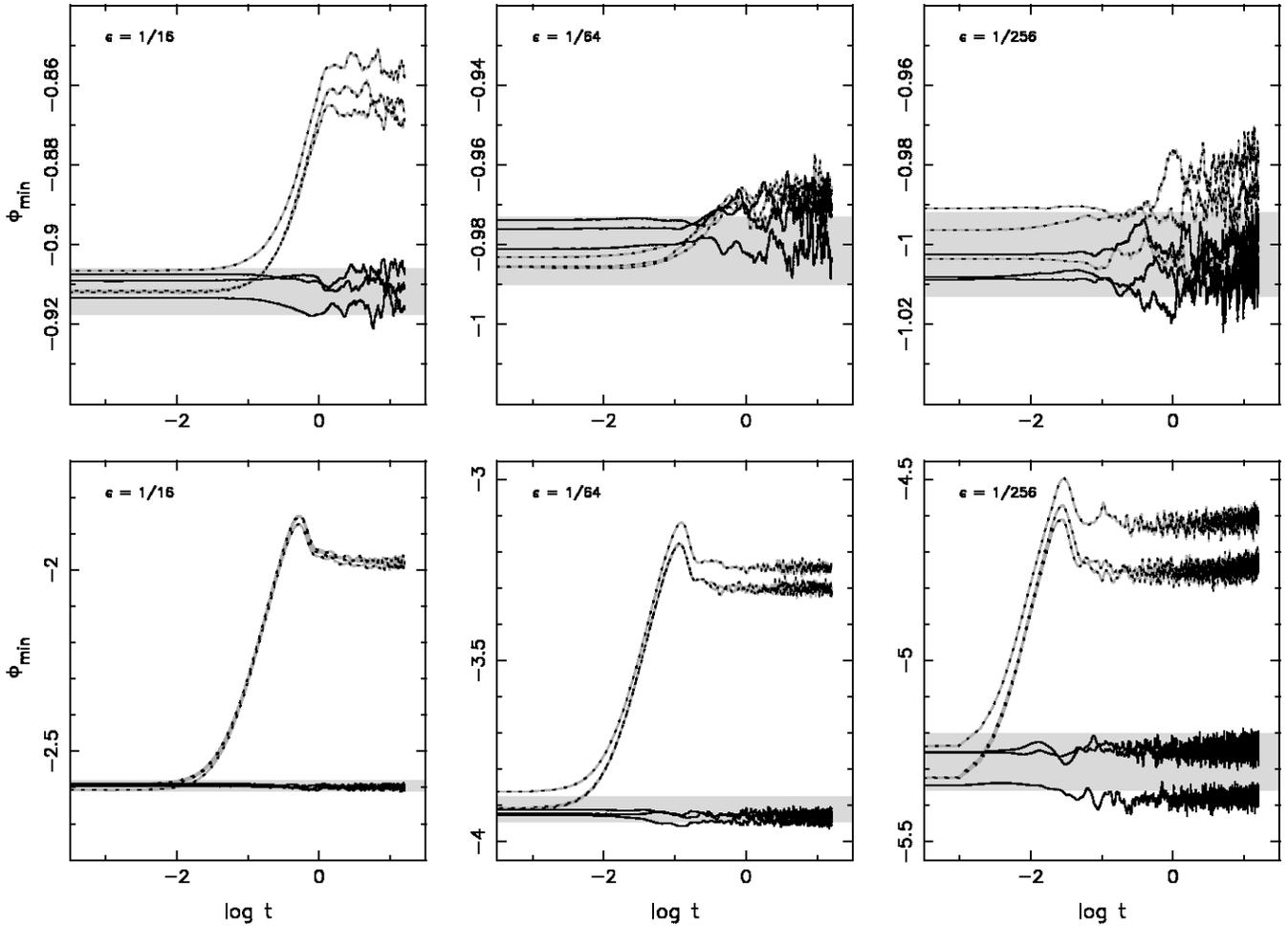}
    \caption{Evolution of potential well depth for $N$-body
    \chng{simulations} of Hernquist (top row) and Jaffe (bottom row)
    models, \chng{each run with the} $\epsilon$ value labeled.
    Solid (\chng{dotted}) curves show results for initial conditions
    generated from \chng{smoothed (self-consistent)} DFs.
    Light grey bands show expected $\pm 2 \sigma$ variation in central
    potential for $N = 262144$ independent particles.
    \label{phiEvol}}
  \end{center}
\end{figure}

Fig.~\ref{phiEvol} shows how the potential \chng{well} depth
$\Phi_\mathrm{min}$ of each simulation evolves as a function of time.
Here, \chng{well} depth is estimated by computing the softened
gravitational potential $\Phi_i$ of each particle $i$ and taking the
minimum (most negative) value.
(This is more accurate than evaluating the potential at the origin
since the center of the system may wander slightly during a
dynamical simulation.)
To better display the observed changes in $\Phi_\mathrm{min}$, the
horizontal axis is logarithmic in time.

Most of the ensembles set up without allowing for softening
(\chng{dotted} curves in Fig.~\ref{phiEvol}) are clearly not in
equilibrium.
In all three of the Jaffe models (bottom row), the potential
wells become dramatically shallower on a time-scale comparable to the
dynamical time at $r = \epsilon$.
The reason for this is evident.
The \chng{self-consistent} Jaffe model has a central potential
diverging like $\log r$ as $r \to 0$; this potential can confine
particles with finite velocity dispersion at arbitrarily small
radii.
However, the relatively shallow potential well of a smoothed Jaffe
model cannot confine these particles; they travel outward in a
coherent surge and phase-mix at radii of a few $\epsilon$.
Their outward surge and subsequent fallback accounts for the rapid
rise and partial rebound of the central potential.
Similar although less pronounced evolution occurs in the Hernquist
models (top row) with $\epsilon = 1/16$ and possibly with $\epsilon =
1/64$ as well.
Only \chng{the self-consistent} Hernquist models run with $\epsilon =
1/256$ \chng{appear truly close to} equilibrium.

In contrast, \chng{all of} the ensembles set up with \chng{smoothed
DFs} (solid curves in Fig.~\ref{phiEvol}) are close to dynamical
equilibrium. 
\chng{In equilibrium, gravitational} potentials fluctuate as
individual particles move along their orbits. 
If particles are uncorrelated, the amplitude of \chng{these}
fluctuations should be comparable to the amplitude seen
\chng{in an ensemble} of independent realizations.
To check this, $1000$ realizations of each model were generated;
\chng{central} gravitational potentials $\Phi_\mathrm{min}$ were
evaluated using the same tree algorithm and parameters used for the
self-consistent simulations.
The grey horizontal bands show a range of $\pm 2 \sigma$ around the
average central potential for each model and choice of $\epsilon$.
Some of the simulations set up \chng{using smoothed DFs} wander
slightly beyond the $2 \sigma$ range.
However, none of them exhibit the dramatic evolution seen in the cases
set up \chng{using self-consistent DFs}.

\begin{figure}
  \begin{center}
    \includegraphics[clip=true,width=\columnwidth,angle=0]{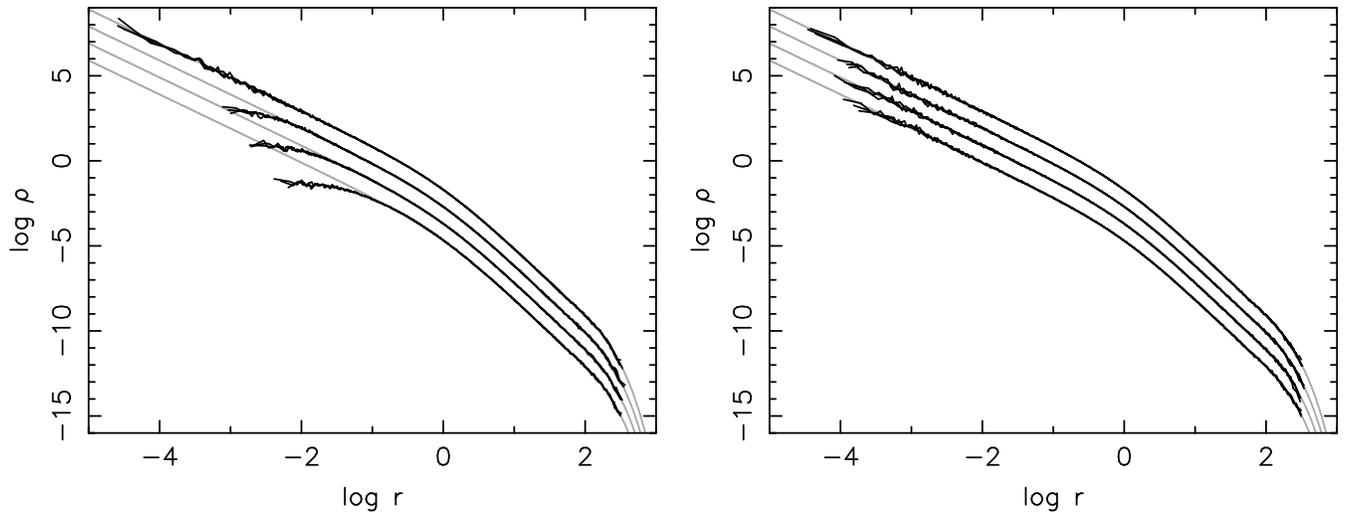}
    \caption{Density profiles of Jaffe models before and after
    dynamical evolution.
    Initial conditions were constructed using \chng{self-consistent
    DFs} (left) and smoothed \chng{DFs} (right).
    In each panel, the top profile shows the initial conditions; the
    smooth curve is the underlying tapered model, overplotted by three
    slightly bumpy curves showing numerical results from three
    independent realizations.
    The three profiles below show numerical results at $t = 16$ for
    simulations run with $\epsilon = 1/256$, $1/64$, and $1/16$,
    respectively; each is displaced downward by one additional unit in
    $\log \rho$ for clarity.
    \label{rhoProfJ}}
  \end{center}
\end{figure}

\chng{The central potential is relatively insensitive to changes in
the mass distribution on scales $r \ll \epsilon$.  To examine
small-scale changes directly}, density profiles measured from the
initial conditions were compared to profiles measured \chng{at $t =
16$} time units.
These profiles were derived as follows.
First, SPH-style \chng{interpolation} with an adaptive kernel
containing $32$ particles was used to estimate the density
\chng{around} each particle.
Next, the centroid \chng{position $\vect{r}_\mathrm{cent}$} of the
$32$ highest-density particles was determined.
Finally, a set of nested spherical shells, \chng{centered on
$\vect{r}_\mathrm{cent}$}, were constructed; each shell \chng{was
required to contain} at least $16$ particles and \chng{have} an outer
radius at least $1.05$ times its inner radius.

\chng{Fig.~\ref{rhoProfJ} summarizes results for Jaffe models, which
display the most obvious changes.
The density of each shell is plotted
against the average distance from $\vect{r}_\mathrm{cent}$ of the
particles it contains.}
In each panel, the top set of curves compare \chng{initial ($t = 0$)}
numerical results with the \chng{underlying} tapered Jaffe model,
\chng{always} represented by \chng{a} light grey line.
\chng{Profiles from three independent $N$-body} realizations of each
model are overplotted.
While \chng{some scatter from realization to realization is seen},
the measured densities \chng{track the underlying profile}
throughout the entire range plotted.
The outermost point is at $\sim 10^7$ times the radius of the
innermost one; there are not enough particles to obtain measurements
\chng{at smaller or larger radii}.

Ranged below the top curves in Fig.~\ref{rhoProfJ} are numerical
results at \chng{$t = 16$} for softening lengths $\epsilon = 1/256$, $1/64$,
and $1/16$, each shifted downward by one more unit in $\log \rho$.
\chng{Again}, profiles from three independent simulations are
overplotted \chng{to illustrate run-to-run variations}.
Simulations set up \chng{using the self-consistent DF} (left panel)
show significant \chng{density evolution}; their initial power-law
profiles are \chng{rapidly} replaced by cores of roughly constant
density \chng{inside} $r \la \epsilon$.
In contrast, simulations set up \chng{using smoothed DFs} (right
panel) follow the initial profile down to $r \sim 10^{-4}$
\chng{(although density evolution occurs on smaller scales).
This shows that a careful set-up procedure can maintain the initial
density profile even on scales much smaller than the softening
length.}

A similar plot for the Hernquist models confirms that most of these
simulations \chng{start} close to equilibrium.
Hernquist models set up \chng{using smoothed DFs} don't appear to
\chng{evolve} at all, \chng{although this statement should be
qualified since the profiles of these models can't be measured
reliably on scales much smaller than $r \sim 10^{-2}$}.
\chng{Models} set up \chng{using the self-consistent DF} and run
with $\epsilon = 1/16$ undergo some \chng{density} evolution; their
profiles \chng{fall below} the underlying Hernquist model at \chng{$r
\sim \epsilon$}, although they continue rising to the innermost point
measured. Simulations run with $\epsilon = 1/64$ or less display no
obvious changes \chng{down to scales of $r \sim 10^{-2}$}.

\chng{It appears} that the \chng{Jaffe} models set up \chng{using
smoothed DFs} are not \textit{completely} free of long-term
evolution.
\chng{The right-hand panel of Fig.~\ref{rhoProfJ} shows that the peak
density as measured using a fixed number of particles falls by roughly
an order of magnitude by $t = 16$.}
\chng{Moreover, a close examination of Fig.~\ref{phiEvol} turns some
cases with a gradual decrease in potential well depth; in the smoothed
Jaffe models with $\epsilon = 1/256$, for example, $\Phi_\mathrm{min}$
exhibits an upward trend of $\sim 0.011$ percent per unit time.}
This evolution \chng{may not be} due to \chng{any} flaw in the initial
conditions; the central \chng{cusps of such models}, which are
confined by \textit{very} shallow harmonic \chng{potentials, are
fragile and easily disrupted.}
\chng{There may be more than one mechanism at work here; the rate of
potential evolution appears to be inversely proportional to particle
number $N$, while the rate of density evolution is independent of $N$.
A full examination of this matter is beyond the scope of this paper}.

\subsubsection{Choice of time-step}

Selecting the time-step $\Delta t$ for an $N$-body simulation is a
non-trivial problem.
While the choice can usually be justified post-hoc by testing for
convergence in a series of experiments with different time-steps, it's
clearly convenient to be able to estimate an appropriate $\Delta t$ a
priori.
A general rule governing such estimates is that the time-step should be
smaller than the shortest dynamical time-scale present in the
simulation.

The central density $\rho_\mathrm{c} = \rho(0;\epsilon)$ of a smoothed
density profile defines one such time-scale.
Within the nearly constant-density core of a smoothed profile, the
local orbital period is $t_\mathrm{c} = \sqrt{3 \pi / G
\rho_\mathrm{c}}$; this is the shortest orbital period anywhere in the
system.
Numerical tests show the leap-frog integrator is well-behaved if
$\Delta t \la 0.05 t_\mathrm{c}$ (conversely, it becomes unstable if
$\Delta t \ga 0.15 t_\mathrm{c}$).
Among the models simulated here, the Jaffe model with $\epsilon =
1/256$ has the highest smoothed central density; for this model,
$\rho_\mathrm{c} = \rho_\mathrm{J}(0;\epsilon) \simeq D_0(2)
\rho_\mathrm{J}(\epsilon) = 10440$.
Given this density, $t_\mathrm{c} \simeq 0.0300$ and $\Delta t \la
0.0015$.

The time required for a fast-moving particle to cross the core region
defines another time-scale.
If $\Phi_\mathrm{c} = \Phi(0;\epsilon)$ is depth of the central
potential well, the maximum speed of a bound particle is $\sqrt{-2
\Phi_\mathrm{c}}$, and the core crossing time is $t_\mathrm{x} =
\epsilon / \sqrt{-2 \Phi_\mathrm{c}}$.
The smoothed Jaffe model with $\epsilon = 1/256$ has the deepest
potential well.
For this model, tests of the leap-frog with fast-moving particles on
radial orbits show that $\Delta t \le t_\mathrm{x} \simeq 0.0012$
yields good results, but time-steps a few times longer result in poor
energy conservation as particles traverse the core region.
(The relationship between $t_\mathrm{x}$ and the maximum acceptable
time-step $\Delta t$ may be somewhat model-dependent.)

Thus, for the Jaffe model with $\epsilon = 1/256$, both the local
criterion based on $t_\mathrm{c}$ and the global criterion based on
$t_\mathrm{x}$ yield similar constraints\footnote{Assuming $\epsilon
\ll a$, one can show that $t_\mathrm{c} / t_\mathrm{x} \simeq 11
\sqrt{\ln (a/\epsilon)}$ for any smoothed Jaffe model; both criteria
yield similar constraints almost independent of $\epsilon$.
For smoothed Hernquist models, on the other hand, $t_\mathrm{c} /
t_\mathrm{x} \simeq 11 \sqrt{a/\epsilon}$; the constraint based on
$t_\mathrm{x}$ is generally stricter.} on $\Delta t$.
It's convenient to round $\Delta t$ down to the next power of two,
implying $\Delta t = 1/1024$.
This corresponds to $\Delta t \simeq 0.03 t_\mathrm{c} \simeq 0.81
t_\mathrm{x}$, which is somewhat conservative but helps insure that
non-equilibrium changes will be accurately followed.

To see if this time-step is reasonable, realizations of this model were
simulated with various values of $\Delta t$ between $1/128$ to
$1/2048$.
At the lower end of this range, the effects of an over-large time-step
manifest quickly; global energy conservation is violated, and the
measured central potential $\Phi_\mathrm{min}$ drifts upward
over time (even though the initial conditions, generated from
$f(E;\epsilon)$, are near equilibrium).
With a time-step $\Delta t = 1/128$, for example; the potential well
becomes $\sim 3$~percent shallower during the first two time-steps, and
by $t = 4$ its depth has decreased by $18$~percent.
These simulations also violate global energy conservation, becoming
$\sim 4.5$~percent less bound by $t = 4$.
Integration errors are reduced -- but not entirely eliminated -- with
a time-step $\Delta t = 1/256$; by $t = 4$, the potential well becomes
$\sim 1.3$~percent shallower, while total energy changes by $\sim
0.4$~percent.
With a time-step of $\Delta t = 1/512$ or less, global energy
conservation is essentially perfect, and variations in
$\Phi_\mathrm{min}$ appear to be driven largely by particle
discreteness as opposed to time-step effects.

Plots analogous to Fig.~\ref{rhoProfJ} show that the simulations with
time-steps as large as $\Delta t = 1/256 \simeq 0.12 t_\mathrm{c}$
reproduce the inner cusps of Jaffe models just as well as those with
$\Delta t = 1/1024$.
With \chng{this time-step}, individual particles may not be followed
accurately, but their aggregate distribution is not obviously
incorrect.
On the other hand, a time-step $\Delta t = 1/128$ yields density
profiles which fall below the initial curves for $r \la 0.01$.

\section{DISCUSSION}

Softening and smoothing are mathematically equivalent.
While the particular form of softening adopted in (\ref{eq:nbody})
corresponds to smoothing with a Plummer kernel
(\ref{eq:plummer-kernel}), other softening prescriptions can also be
described in terms of smoothing operations \citep[e.g.][]{D01}.
There are two conceptual advantages to thinking about softening as a
smoothing process \chng{transforming} the underlying density field
$\rho(\vect{r})$ to the smoothed density field
$\rho(\vect{r};\epsilon)$.
First, since the gravitational potential $\Phi(\vect{r};\epsilon)$ is
related to $\rho(\vect{r};\epsilon)$ by Poisson's equation, the
powerful mathematical machinery of classical potential theory becomes
available to analyze \chng{potentials in} $N$-body simulations.
Second, focusing attention on smoothing makes the source term for the
gravitational field explicit.
\chng{From this perspective}, smoothing is not a property of the
particles, but a separate step introduced to ameliorate $1/r$
singularities in the potential.
Particle themselves are points \chng{rather than} extended objects;
this insures that their trajectories are
characteristics of (\ref{eq:vlasov}).

Plummer smoothing converts $\rho_n(r) \propto r^{-n}$ power-laws to
cores.
At radii $r \gg \epsilon$ the density profile is essentially
unchanged, while at $r \ll \epsilon$ the density approaches a constant
value equal to the density of \chng{the} underlying model at $r =
\epsilon$ times a factor which depends only on $n$.
For the case $n = 1$, this factor is unity and $\rho_1(r;\epsilon) =
\rho_1(\sqrt{r^2 + \epsilon^2})$ everywhere.

The effects of Plummer smoothing on astrophysically-motivated models
with power-law cusps, such as the Jaffe, Hernquist, and NFW profiles,
follow for the most part from the results for pure power-laws.
In particular, for $\epsilon \la a/64$, where $a$ is the profile's
scale radius, the power-law results are essentially `grafted' onto the
underlying profile.
On the other hand, for $\epsilon \ga a/16$, the inner power-law is
erased by smoothing.

Smoothing provides a way to predict the potentials obtained in
$N$-body calculations to an accuracy limited only by $\sqrt{N}$
fluctuations.
These predictions offer new and powerful tests of $N$-body
methodology, exposing subtle systematic effects which may be difficult
to diagnose by other means.

Given an underlying density profile $\rho(r)$, it's straightforward to
construct an isotropic distribution function $f(E;\epsilon)$ such that
$\rho(r)$ is in equilibrium with the potential generated by its
smoothed counterpart $\rho(r;\epsilon)$.
Such distribution functions \chng{can be used} to generate
\chng{high-quality} equilibrium initial conditions for $N$-body
simulations; \chng{they should be particularly effective when realized
with `quiet start' procedures \citep{DS00}}.
Systems with shallow central cusps, such as Hernquist and NFW models,
may be set up fairly close to equilibrium without \chng{taking}
softening \chng{into account} as long as  $\epsilon$ is \chng{not too
large}.
However, it appears impossible to set up a good $N$-body realization
of a Jaffe model without allowing for softening.

It's true that realizations so constructed \chng{don't} reproduce the
actual dynamics of the underlying models at small radii
\citep[][footnote 8]{D01}; to \chng{obtain an} equilibrium, the
velocity dispersion \chng{is reduced} on scales $r \la \epsilon$.
But realizations set up \textit{without} softening preserve neither
the dispersion nor the density at small radii, and the initial
relaxation of such a system can't be calculated a priori but must be
simulated numerically.
On the whole, it seems better to get the central density profile right
on scales $r < \epsilon$, and know how the central velocity dispersion
profile has been modified.
Even if the dynamics are not believable \chng{within} $r \la
\epsilon$, the ability to localize mass on such scales may be
advantageous in modeling dynamics on larger scales.

Mathematica code to tabulate smoothed models is available at\\
\texttt{http://www.ifa.hawaii.edu/faculty/barnes/research/smoothing/}.

\section*{ACKNOWLEDGMENTS}
 
I thank Jun Makino and Lars Hernquist for useful and encouraging
comments, \chng{and an anonymous referee for a positive and
constructive report}.  Mathematica rocks.

\section*{APPENDIX A: APPROXIMATIONS}

\subsection*{A.1 Power-law profiles}

The power-law density and cumulative mass profiles are
\begin{equation}
  \rho_n(r) = \rho_\mathrm{a} \left( \frac{a}{r} \right)^n \, ,
  \qquad
  M_n(r) = \frac{4 \pi}{3 - n} \rho_\mathrm{a} a^n r^{(3 - n)} \, .
\end{equation}

Plummer smoothing converts power laws with $n < 3$ to finite-density
cores.
At $r \ll \epsilon$ the smoothed density is nearly constant and close
to the smoothed central density $\rho_n(0;\epsilon) = D_0(n)
\rho_n(\epsilon)$.
Within this constant-density region, the smoothed mass profile is
approximately
\begin{equation}
  \overline{M}_n(r;\epsilon) =
    \frac{4 \pi}{3} r^3 \rho_n(0;\epsilon) =
    \frac{4 \pi}{3} r^3
      D_0(n) \rho_\mathrm{a} \left( \frac{a}{\epsilon} \right)^n \, .
\end{equation}
At $r \gg \epsilon$, on the other hand, smoothing has little effect on
the mass profile, so $M_n(r;\epsilon) \simeq M_n(r)$.
Interpolating between these functions yields an approximate expression
for the smoothed mass profile:
\begin{equation}
  \widetilde{M}_n(r;\epsilon) =
    \left(\overline{M}_n(r;\epsilon)^{-\kappa/n} +
          M_n(r)^{-\kappa/n}\right)^{-n/\kappa} \, ,
\end{equation}
where the shape parameter $\kappa$ determines how abruptly the
transition from one function to the other takes place.
This expression can be rearranged to give
\begin{equation}
  \widetilde{M}_n(r;\epsilon) =
    \left( \left(\frac{3}{(3 - n) D_0(n)}\right)^{\kappa/n}
           \left(\frac{\epsilon}{r}\right)^\kappa + 1 \right)^{-n/\kappa} \,
      M_n(r) \, .
  \label{eq:approx-smooth-mass}
\end{equation}
The smoothed density profile is obtained by differentiating the mass
profile:
\begin{equation}
  \tilde{\rho}_n(r;\epsilon) =
    \frac{1}{4 \pi r^2} \frac{d}{dr} \widetilde{M}_n(r;\epsilon) \, .
  \label{eq:approx-smooth-rho}
\end{equation}

\begin{figure}
  \begin{minipage}{0.45\columnwidth}
    \begin{center}
    \includegraphics[clip=true,width=\columnwidth]{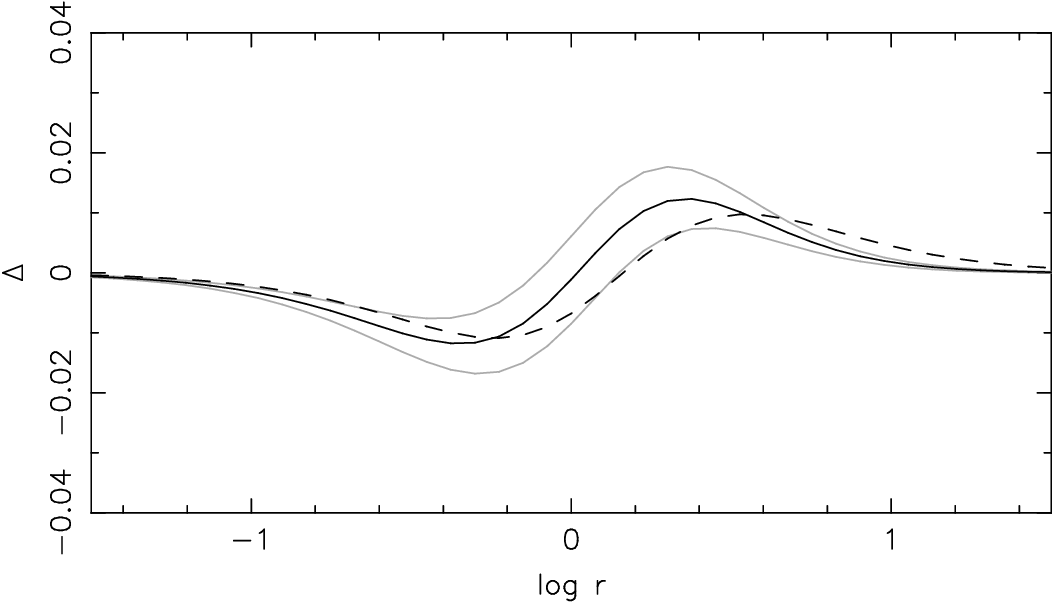}
    \caption{Relative error in smoothed density (solid) and mass
    (dashed) for a $\rho \propto r^{-1}$ profile, computed for
    $\epsilon = 1$ using (\ref{eq:approx-smooth-mass}) and
    (\ref{eq:approx-smooth-rho}).
    Dark curves show results for $\kappa = 1.739$; light grey solid
    curves show errors in density only for $\kappa = 1.769$ (above)
    and $\kappa = 1.709$ (below).
    \label{errorCusp1}}
    \end{center}
  \end{minipage}%
  \hbox to 0.10\columnwidth{}%
  \begin{minipage}{0.45\columnwidth}
    \begin{center}
    \includegraphics[clip=true,width=\columnwidth]{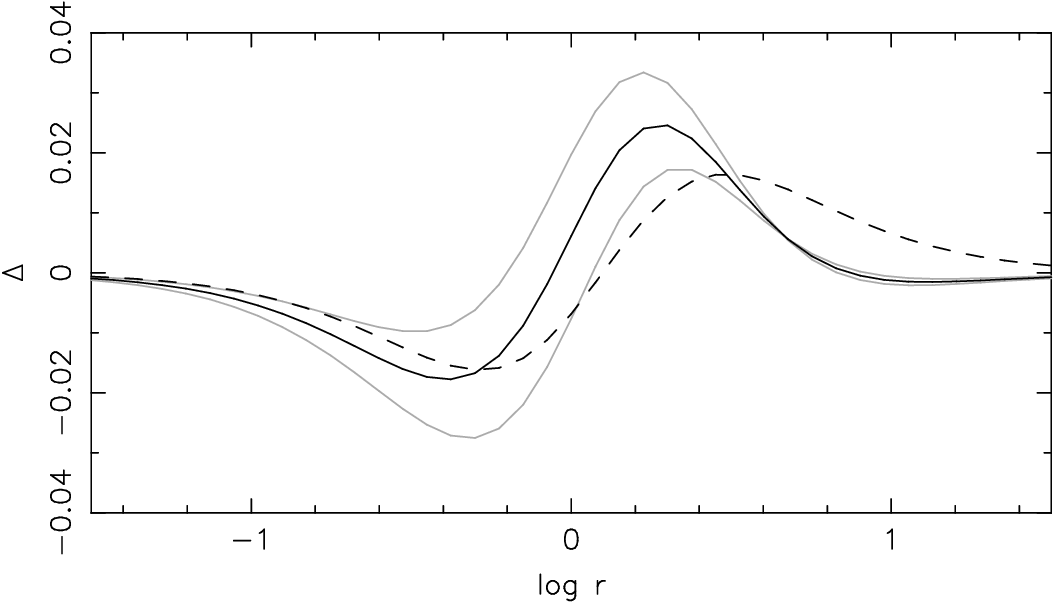}
    \caption{Relative error in smoothed density (solid) and mass
    (dashed) for a $\rho \propto r^{-2}$ profile, computed for
    $\epsilon = 1$ using (\ref{eq:approx-smooth-mass}) and
    (\ref{eq:approx-smooth-rho}).
    Dark curves show results for $\kappa = 1.820$; light grey solid
    curves show errors in density only for $\kappa = 1.850$ (above)
    and $\kappa = 1.790$ (below).
    \label{errorCusp2}}
    \end{center}
  \end{minipage}%
\end{figure}

Figs.~\ref{errorCusp1} and~\ref{errorCusp2} present tests of these
approximations for $\rho \propto r^{-1}$ and~$r^{-2}$ power-laws,
respectively.
As in Figs.~\ref{cusp1} and~\ref{cusp2}, the smoothed density profile
was computed with $\epsilon = 1$; for other values of $\epsilon$, the
entire pattern simply shifts left or right without changing shape or
amplitude.
Dashed curves show relative errors in smoothed mass, $\Delta_M =
\widetilde{M}(r;\epsilon) / M(r;\epsilon) - 1$, while solid curves are
relative errors in smoothed density $\Delta_\rho =
\tilde{\rho}(r;\epsilon) / \rho(r;\epsilon) - 1$.
The $\kappa$ value used for each dark curve is the value which
minimizes $\sum_i \Delta_\rho(r_i)^2$ evaluated at points $r_i$
distributed uniformly in $\log r$ between $\log r = -1.5$ and~$1.5$.
In light grey, plots of $\Delta_\rho(r)$ for two other $\kappa$
illustrate the sensitivity to this parameter.
Comparing these plots, it appears that the approximation works better
for the $\rho \propto r^{-1}$ power-law than it does for $r^{-2}$, but
even in the latter case the maximum error is only $\sim 2$\%.

Because (\ref{eq:approx-smooth-mass}) modifies the underlying mass
profile with a multiplicative factor, it can also be used to
approximate effects of softening on non-power-law profiles
\citep[e.g.][]{BH09}; for this purpose, both $\epsilon$ and $\kappa$
can be treated as free parameters and adjusted to provide a good fit.
The resulting errors in density, which amount to a few percent near
the softening scale, are undesirable but don't \chng{appear to}
seriously compromise $N$-body simulations with $N \sim 10^5$. 

\subsection*{A.2 Hernquist and NFW profiles}

\begin{figure}
  \begin{center}
    \includegraphics[clip=true,width=0.45\columnwidth]{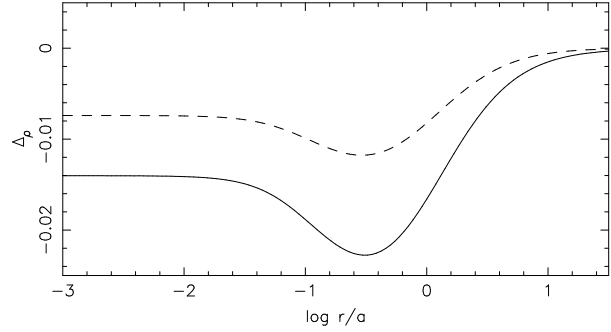}
    \caption{Relative error in density, $\Delta_\rho = \tilde{\rho} /
    \rho - 1$, plotted vs.~radius $r$, for the approximations given in
    (\ref{eq:approx-rho-cusp1}).
    Solid and dashed curves show results for Hernquist and NFW
    profiles, respectively, computed for $\epsilon = a/16$.
    \label{errorRhoHNFW}}
  \end{center}
\end{figure}

For $\epsilon \ll a$, \chng{smoothing} primarily modifies the $r^{-1}$
part of \chng{these} density \chng{profiles}.
This, together with the exact solution for the case $\rho \propto
r^{-1}$ given \chng{in \S~2.1}, suggests simple approximations for
smoothed Hernquist and NFW models:
\begin{equation}
  \tilde{\rho}_\mathrm{H}(r;\epsilon) =
    \rho_\mathrm{H}({\textstyle \sqrt{r^2 + \epsilon^2}}) =
      \rho_\mathrm{H}(r_\epsilon) \, ,
  \qquad
  \tilde{\rho}_\mathrm{NFW}(r;\epsilon) =
    \rho_\mathrm{NFW}({\textstyle \sqrt{r^2 + \epsilon^2}}) =
      \rho_\mathrm{NFW}(r_\epsilon) \, .
  \label{eq:approx-rho-cusp1}
\end{equation}
Fig.~\ref{errorRhoHNFW} plots the relative error in density,
$\Delta_\rho = \tilde{\rho}(r;\epsilon) / \rho(r;\epsilon) - 1$ for
both models, adopting $\epsilon = a/16$.
For other values of $\epsilon$, these errors scale roughly as
$\epsilon^{1.6}$.

The general behavior of these approximations is readily understood.
Overall, \chng{$\tilde{\rho}_\mathrm{NFW}(r;\epsilon)$ is more
accurate than $\tilde{\rho}_\mathrm{H}(r;\epsilon)$ since the NFW
profile} is closer to $\rho \propto r^{-1}$ at all radii.
Both curves are approximately flat for $r \ll \epsilon$, then
\chng{reach minima} for $r$ between $\epsilon$ and the profile scale
\chng{radius} $a$.
These \chng{minima} arise because the smoothed density \chng{approaches
or even} slightly exceeds the underlying density (see Figs.~\ref{rhoH}
and \ref{rhoNFW}), while \chng{(\ref{eq:approx-rho-cusp1})} always
\chng{yields values below} the underlying density.

It's sometimes useful to have the cumulative mass for a smoothed
profile.
The approximate profiles in (\ref{eq:approx-rho-cusp1}) can be
integrated analytically, although the resulting expressions are a bit
awkward:
%%%%%%%%%%%%%%%%%%%%%%%%%%%%%%%%%%%%%%%%%%%%%%%%%%%%%%%%%%%%%%%%%%%%%%%%
\begin{eqnarray}
  \widetilde{M}_\mathrm{H}(r;\epsilon) &=&
    \int_{0}^{r} dx \, 4 \pi x^2 \, \tilde{\rho}_\mathrm{H}(x;\epsilon) \\
                                       &=& \textstyle 
    \frac{\epsilon^2 a M}{(\epsilon^2-a^2)^{3/2}}
      \left(\arctan\left(\frac{r}{\sqrt{\epsilon^2-a^2}}\right) -
            \arctan\left(\frac{a r}{r_\epsilon \sqrt{\epsilon^2-a^2}}\right) -
            \frac{r \sqrt{\epsilon^2 - a^2}
                    \left(a^3 r_\epsilon - a^2 (\epsilon^2 + 2r^2) +
                          a r_\epsilon (r^2 - \epsilon^2) +
                          \epsilon^2 r_\epsilon^2\right)}
                 {\epsilon^2 (r_\epsilon^2 - a^2)^2}\right)
    \nonumber
\end{eqnarray}
%%% \begin{eqnarray}
%%%   \textstyle
%%%   \widetilde{M}_\mathrm{H}(r;\epsilon) &=&
%%%     \int_{0}^{r} dx \, 4 \pi x^2 \, \tilde{\rho}_\mathrm{H}(x;\epsilon) \\
%%%                                        &=& \scriptstyle 
%%%     \frac{\epsilon^2 a M}{(\epsilon^2-a^2)^{3/2}}
%%%       \left(\arctan\left(\frac{r}{\sqrt{\epsilon^2-a^2}}\right) -
%%%             \arctan\left(\frac{a r}{r_\epsilon \sqrt{\epsilon^2-a^2}}\right) -
%%%             \frac{r \sqrt{\epsilon^2 - a^2}
%%%                     \left(a^3 r_\epsilon - a^2 (\epsilon^2 + 2r^2) +
%%%                           a r_\epsilon (r^2 - \epsilon^2) +
%%%                           \epsilon^2 r_\epsilon^2\right)}
%%%                  {\epsilon^2 (r_\epsilon^2 - a^2)^2}\right)
%%%     \nonumber
%%% \end{eqnarray}
%%%%%%%%%%%%%%%%%%%%%%%%%%%%%%%%%%%%%%%%%%%%%%%%%%%%%%%%%%%%%%%%%%%%%%%%
and
%%%%%%%%%%%%%%%%%%%%%%%%%%%%%%%%%%%%%%%%%%%%%%%%%%%%%%%%%%%%%%%%%%%%%%%%
\begin{eqnarray}
  \widetilde{M}_\mathrm{NFW}(r;\epsilon) &=&
    \int_{0}^{r} dx \, 4 \pi x^2 \, \tilde{\rho}_\mathrm{NFW}(x;\epsilon) \\
                                         &=& \textstyle
    4 \pi a^3 \rho_\mathrm{a}
      \left(\frac{r (a - r_\epsilon)}{r_\epsilon^2 - a^2} +
	    \frac{a}{\sqrt{\epsilon^2-a^2}}
	      \left(\arctan\left(\frac{a r}{r_\epsilon
					      \sqrt{\epsilon^2-a^2}}\right) -
		    \arctan\left(\frac{r}
                                      {\sqrt{\epsilon^2-a^2}}\right)\right) +
	    \log\left(\frac{r + r_\epsilon}{\epsilon}\right)\right)
    \nonumber
\end{eqnarray}
%%% \begin{eqnarray}
%%%   \textstyle
%%%   \widetilde{M}_\mathrm{NFW}(r;\epsilon) &=&
%%%     \int_{0}^{r} dx \, 4 \pi x^2 \, \tilde{\rho}_\mathrm{NFW}(x;\epsilon) \\
%%%                                          &=& \scriptstyle
%%%     4 \pi a^3 \rho_\mathrm{a}
%%%       \left(\frac{r (a - r_\epsilon)}{r_\epsilon^2 - a^2} +
%%% 	    \frac{a}{\sqrt{\epsilon^2-a^2}}
%%% 	      \left(\arctan\left(\frac{a r}{r_\epsilon
%%% 					      \sqrt{\epsilon^2-a^2}}\right) -
%%% 		    \arctan\left(\frac{r}
%%%                                       {\sqrt{\epsilon^2-a^2}}\right)\right) +
%%% 	    \log\left(\frac{r + r_\epsilon}{\epsilon}\right)\right)
%%%     \nonumber
%%% \end{eqnarray}
%%%%%%%%%%%%%%%%%%%%%%%%%%%%%%%%%%%%%%%%%%%%%%%%%%%%%%%%%%%%%%%%%%%%%%%%
Note that because the approximate profiles (\ref{eq:approx-rho-cusp1})
systematically underestimate the true smoothed \chng{densities}, these
expressions will likewise systematically underestimate the total
smoothed mass.

\section*{APPENDIX B: FORCE CALCULATION ERRORS}

Tree codes reduce the computational cost of gravitational force
calculation by making explicit approximations \citep{BH86}.
The long-range potential due to a localized mass distribution
$\mathcal{M}$ with total mass $M$ and center of mass position
$\vect{r}_0$ is approximated as
\begin{equation}
  \Phi(\vect{r}) =
    - \frac{G M}{|\vect{r} - \vect{r}_0|} + \mathrm{higher\ order\ terms} \, ,
\end{equation}
where the higher order terms include quadrupole and possibly
higher-order moments (dipole terms vanish because $\vect{r}_0$
coincides with the center of mass).
To implement softening, this approximation is typically replaced with
\begin{equation}
  \Phi(\vect{r}) =
    - \frac{G M}{\sqrt{|\vect{r} - \vect{r}_0|^2 + \epsilon^2}} +
    \mathrm{higher\ order\ terms} \, .
  \label{eq:approx-pot-soft}
\end{equation}
This works at large distances, but becomes inaccurate if $|\vect{r} -
\vect{r}_0| \sim \epsilon$ \citep{H87}.
Moreover, because the error is introduced at the \textit{monopole}
level \citep{WC06}, higher-order corrections don't repair the damage.

To appreciate the problem, consider a sphere $\mathcal{S}$ centered on
$\vect{r}_0$ with radius $R$ large enough to enclose $\mathcal{M}$.
For $\epsilon = 0$, the inward acceleration averaged over the surface
of $\mathcal{S}$ is easily computed using Gauss's theorem:
\begin{equation}
  \overline{a}_r \equiv
  \frac{1}{4 \pi R^2}
    \int_{\partial\mathcal{S}} d\vect{A} \, \cdot \vect{a} =
  - \frac{G M}{R^2} \, .
\end{equation}
In other words, the monopole term is sufficient to calculate the
inward acceleration averaged over the surface of $\mathcal{S}$
\textit{exactly}.

Suppose we want to compute $\overline{a}_r$ for $\epsilon > 0$.
Again using Gauss's theorem, we have
\begin{equation}
  \overline{a}_r = - \frac{G M_\mathcal{S}(\epsilon)}{R^2} \, ,
  \qquad \mathrm{where} \qquad
  M_\mathcal{S}(\epsilon) =
    \int_\mathcal{S} d\vect{r} \, \rho(\vect{r};\epsilon)
\end{equation}
is the smoothed mass within the sphere.
As before, this is an exact equality.
The tree code approximation (\ref{eq:approx-pot-soft}) implies that
the enclosed mass is
\begin{equation}
  M_\mathcal{S}^0(\epsilon) = \frac{M}{(1 + \epsilon^2/R^2)^{3/2}} \, .
\end{equation}
This is correct if $\mathcal{M}$ is simply a point mass located at
$\vect{r}_0$.
But if $\mathcal{M}$ has finite extent, then the enclosed mass
$M_\mathcal{S}(\epsilon)$ is \textit{always} less than
$M_\mathcal{S}^0(\epsilon)$.
As a result, (\ref{eq:approx-pot-soft}) will systematically
\textit{overestimate} the inward acceleration and depth of the
potential well due to $\mathcal{M}$.
\citet{WC06} demonstrate a similar result by computing the softened
potential of a homogeneous sphere; they find $- G M / \sqrt{|\vect{r}
- \vect{r}_0|^2 + \epsilon^2}$ is only the first term in a series.

The inequality $M_\mathcal{S}(\epsilon) < M_\mathcal{S}^0(\epsilon)$
is easily verified for Plummer softening.
An analogous inequality is likely to hold for other smoothing kernels
$S(r;\epsilon)$ which monotonically decrease with $r$.
Smoothing kernels with compact support \citep{D01} may be better
behaved in this regard.

Under what conditions are these errors significant?  For `reasonable'
values of $\epsilon$, most dynamically relevant interactions are on
ranges $\Delta r \gg \epsilon$ where softening has little effect;
these interactions are not compromised since
(\ref{eq:approx-pot-soft}) is nearly correct at long range.
Only if a significant fraction of a system's mass lies within a region
of size $\epsilon$ can these errors become important.
This situation was not investigated in early tree code tests
\citep[e.g.,][]{H87,BH89}, which generally used mass models with cores
instead of central cusps, and even heavily softened Hernquist models
don't have much mass within one softening radius.
On the other hand, Jaffe models pack more mass into small radii; a
Jaffe model with $\epsilon = a/16$ has almost $6$~percent of its mass
within $r = \epsilon$.
Jaffe models should be good test configurations for examining treecode
softening errors.

Tests were run using tapered Jaffe and Hernquist models, realized
using the same parameters ($a = 1$, $b = 100$, $M = 1$, and $N =
262144$) used in the dynamical experiments (\S~4.3).
In each model, the gravitational field was sampled at $4096$ points
drawn from the same distribution as the mass.
At each test point, results from the tree code with opening angle
$\theta = 0.8$, including quadrupole terms, were compared with the
results of an direct-sum code.
As expected, the tests with Hernquist models showed relatively little
trend of force calculation error with $\epsilon$, although the errors
are somewhat larger for $\epsilon = 1/16$ than for smaller values.
In contrast, the tests with Jaffe models reveal a clear relationship
between softening length and force calculation accuracy.

\begin{figure}
  \begin{minipage}[t]{0.45\columnwidth}
    \includegraphics[clip=true,width=\columnwidth]{accelErrorJ.ps}

    \caption{Tree code acceleration error $\delta a / a$ plotted
    against radius.
    These results were obtained for a Jaffe model with $\theta = 0.8$.
    Grey dots show errors for individual test points with $\epsilon =
    1/64$; the jagged curve threading the dots is constructed by
    averaging points in groups of $16$.
    Similar curves above and below show results for $\epsilon = 1/16$
    and $\epsilon = 1/256$, respectively.
    The large marker on each curve shows the average value of $\delta
    a / a$ at radius $r = \epsilon$.
    The light grey curve shows results for $\epsilon = 0$.
    \label{accelErrorJ}}
  \end{minipage}%
  \hbox to 0.10\columnwidth{}%
  \begin{minipage}[t]{0.45\columnwidth}
    \includegraphics[clip=true,width=\columnwidth]{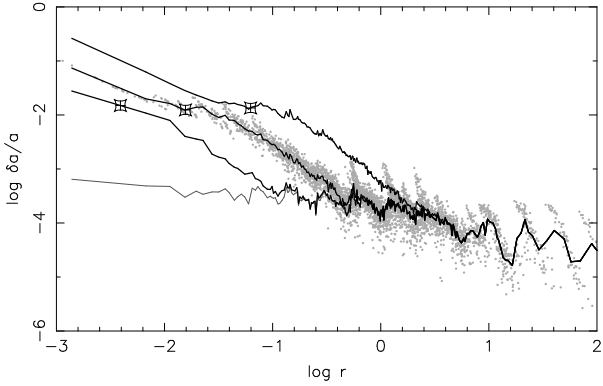}
    \caption{Jaffe model results for evolution of difference in
    potential well depth $\Phi_\mathrm{min}$ for simulations run with
    $\theta = 0.8$ and $0.4$.  
    Light grey, dark grey, and black show potential differences for
    $\epsilon = 1/256$, $1/64$, and $1/16$, respectively; three
    independent realizations are plotted in each case.
    \label{phiDiffJ}}
  \end{minipage}%
\end{figure}

Fig.~\ref{accelErrorJ} shows the relative acceleration error $\delta a
/ a = |\vect{a}_\mathrm{t} - \vect{a}_\mathrm{d}| /
|\vect{a}_\mathrm{d}|$ for Jaffe models with various values of
$\epsilon$.
Here $\vect{a}_\mathrm{t}$ and $\vect{a}_\mathrm{d}$ are accelerations
computed using a tree code and direct summation, respectively.
The grey dots represent measurements of $\delta a / a$ for individual
test points, computed using $\epsilon = 1/64$.
The pattern of errors suggests two regimes.
At radii $r \ga 0.25$ ($\log r \ga -0.4$), the points fall in a
`sawtooth' pattern which reflects the hierarchical cell structure used
in the force calculation.
At smaller radii, on the other hand, the relative error grows more or
less monotonically as $r \to 0$.
It appears that errors in the large-$r$ regime are due to neglect of
moments beyond quadrupole order in computing the potentials of
individual cells; conversely, the errors in the small-$r$ regime are
due to the tree code's inaccurate treatment of softening.
The direction of the error vectors $\vect{a}_\mathrm{t} -
\vect{a}_\mathrm{d}$ supports this interpretation; in the large-$r$
regime they are isotropically distributed, while in the small-$r$
regime they point toward the center of the system.

The jagged line threading through the dots in Fig.~\ref{accelErrorJ},
constructed by averaging test points in groups of $16$, shows the
overall relationship between acceleration error and radius for
$\epsilon = 1/64$.
Similar curves are also plotted for $\epsilon = 1/16$ (above) and
$\epsilon = 1/256$ (below).
At large radii, all three curves coincide precisely, implying that
force errors are independent of $\epsilon$.
Going to smaller $r$, the curve for $\epsilon = 1/16$ is the first to
diverge, rising above the other two, next the curve for $\epsilon =
1/64$ begins to rise, tracking the mean distribution of the plotted
dots, and finally the curve for $\epsilon = 1/256$ parallels the other
two.
Each curve begins rising monotonically at a radius \chng{$r \sim 20
\epsilon$}; this is evidently where softening errors begin to dominate
other errors in the force calculation.
At the softening radius $r = \epsilon$, all three curves show mean
acceleration errors $\delta a / a \simeq 0.013$.

Are errors of this magnitude dynamically important?
In particular, could they explain some of the potential evolution seen
in the runs set up with softening (solid curves) in
Fig.~\ref{phiEvol}?
One possible test is to re-run the simulations using smaller $\theta$
values; decreasing $\theta$ from $0.8$ to $0.6$ or $0.4$ reduces the
tree code acceleration error associated with softening by factors of
$\sim 2$ or $\sim 3$, respectively (albeit at significant
computational costs).
The new runs were started using exactly the same initial conditions as
their $\theta = 0.8$ counterparts.
This initially allows central potentials $\Phi_\mathrm{min}(t)$ to be
compared between runs to high accuracy, temporarily circumventing the
effects of $\sqrt{N}$ fluctuations.

Fig.~\ref{phiDiffJ} compares Jaffe model results for $\theta = 0.8$
and $0.4$.
\chng{Initially}, the potential difference
$\Phi_\mathrm{min}(t;\theta=0.8) - \Phi_\mathrm{min}(t;\theta=0.4)$ 
arises because the treecode systematically overestimates the potential
well depth \chng{by an amount which} is greater for larger values of
$\theta$.
\chng{As the simulations run}, the excess radial
acceleration causes systems run with $\theta = 0.8$ to contract
relative to those run with $\theta = 0.4$, causing a further increase
in potential well depth.
This contraction takes place on a dynamical time-scale at $r \sim
\epsilon$, occurring first for the $\epsilon = 1/256$ simulations
(light grey curves).
At later times, as trajectories in otherwise-identical simulations
with different $\theta$ values diverge, short-term fluctuations in
$\Phi_\mathrm{min}$ de-correlate and the dispersion in potential
differences increases markedly.

\chng{These results show} that force-calculation errors can
have measurable, \chng{albeit modest}, effects on \chng{dynamical
evolution}.
Extrapolating results for a range of $\theta$ values to $\theta = 0$,
it appears that \chng{for $\theta = 0.8$} the \textit{dynamical}
response of these Jaffe models due to excess radial acceleration
deepens central potential wells by \chng{about $0.2$~percent}.
To test this extrapolation, it would be instructive to
repeat these experiments with a direct-summation code.
However, compared to the overall range of $\Phi_\mathrm{min}$
variations seen in Fig.~\ref{phiEvol}, the perturbations due to force
calculation errors seem relatively insignificant.
In addition, density profiles measured from runs with $\theta = 0.6$
or $0.4$ appear \chng{similar to} those shown in Fig.~\ref{rhoProfJ},
again implying that treecode force calculation errors have little
effect on the key results of this study.

\chng{The best way to correct (\ref{eq:approx-pot-soft}) for
short-range interactions is not obvious.
A simple, ad-hoc option is to reduce the effective opening angle on
small scales.
For example, accepting cells which satisfy
\begin{equation}
  d > \frac{\ell}{\theta_\mathrm{eff}} + \delta
  \qquad \mathrm{where} \qquad
  \theta_\mathrm{eff} =
    \theta \, \frac{\epsilon + \ell}{2\epsilon + \ell} \, ,
\end{equation}
where $d$ is the distance to the cell's center of mass, $\ell$ is the
cell's size, measured along any edge, and $\delta$ is the distance
between the cell's center of mass and its geometric center
\citep{B98},  reduces softening-relating treecode errors by a factor
of $\sim 2$ at a modest cost in computing time.
Further experiments with similar expressions may produce better
compromises between speed and accuracy.}

\end{document}